\documentclass[12pt]{article}

\usepackage{amsmath,amssymb,amsthm}
\usepackage{latexsym,graphicx,color,subfigure,slashed}

\newcommand{\beq}{\begin{equation}}
\newcommand{\eeq}[1]{\label{#1}\end{equation}}
\def\beqa{\begin{eqnarray}}
\def\eeqa#1{\label{#1}\end{eqnarray}}
\newcommand{\eeqn}{\end{equation}}
\newcommand{\CR}{\notag \\}
\newcommand{\leqn}[1]{(\ref{#1})}                                                                        

\def\stacksymbols #1#2#3#4{\def\theguybelow{#2}
    \def\vp{\lower#3pt}
    \def\sp{\baselineskip0pt\lineskip#4pt}
    \mathrel{\mathpalette\intermediary#1}}

\def\intermediary#1#2{\vp\vbox{\sp
     \everycr={}\tabskip0pt
     \halign{$\mathsurround0pt#1\hfil##\hfil$\crcr#2\crcr
              \theguybelow\crcr}}}

\begin{document}

\begin{titlepage}

\begin{center}
 {\Huge \bf Higgs Couplings and Naturalness} 
\end{center}

\begin{center}
{\bf Marco Farina, Maxim Perelstein and Nicolas Rey-Le Lorier} \\
\end{center}
\vskip 8pt

\begin{center}
	{\it Laboratory of Elementary Particle Physics, \\
	     Physical Sciences Building\\
	     Cornell University, Ithaca, NY 14853, USA } \\

\vspace*{0.3cm}

{\tt  mf627@cornell.edu, mp325@cornell.edu, nr323@cornell.edu}
\end{center}

\vglue 0.3truecm

\begin{abstract}
\noindent Many extensions of the Standard Model postulate the existence of new weakly coupled particles, the top partners, at or below the TeV scale. The role of the top partners is to cancel the quadratic divergence in the Higgs mass parameter due to top loops. We point out the generic correlation between naturalness (the degree of fine-tuning required to obtain the observed electroweak scale), and the size of top partner loop contributions to Higgs couplings to photons and gluons. If the fine-tuning is required to be at or below a certain level, a model-independent lower bound on the deviations of these Higgs couplings from the Standard Model can be placed (assuming no cancellations between contributions from various sources). Conversely, if a precise measurement of the Higgs couplings shows no deviation from the Standard Model, a certain amount of fine-tuning would be required. We quantify this connection, and argue that a measurement of the Higgs couplings at the per-cent level would provide a serious and robust test of naturalness. 

\end{abstract}

\end{titlepage}

\section{Introduction}
The recent discovery of a new particle, roughly consistent with the Standard Model (SM) Higgs boson, has opened a new window into physics at the electroweak scale. In the next decade, the Higgs physics will enter the precision era, in which the goal will be to measure the properties of this particle, in particular its couplings, with the highest possible accuracy. Besides the continuing experiments at the LHC, the idea of a next-generation electron-positron collider such as the International Linear Collider (ILC) is currently under active discussion, with precise measurements of the Higgs couplings as its prime motivation~\cite{Peskin}. Such a facility would be capable of measuring several couplings  
at a per cent level. It is important to understand the implications that these measurements could have on our ideas about physics beyond the Standard Model. 

Predictions of many SM extensions for the Higgs couplings have already been extensively studied. In this paper, we point out a very general, and important, feature of such predictions. In any model which stabilizes the Higgs mass against radiative corrections by postulating weakly-coupled new physics, the amount of fine-tuning required to obtain the observed electroweak scale is {\it inversely correlated} with the size of certain non-SM contributions to the Higgs couplings to photons and gluons.  In other words, if the fine-tuning is required to be at or below a certain level, a model-independent {\it lower bound} on the deviations of these Higgs couplings from the SM can be placed (assuming no cancellations between contributions from various sources). Conversely, if a precise measurement of the Higgs couplings shows no deviation from the SM, a certain amount of fine-tuning would be required. We will quantify these statements, and show that per-cent level Higgs coupling measurements, expected to be achievable at the next-generation experimental facilities, would provide a serious test of naturalness of the electroweak scale. This gives a clear and compelling physics motivation for such measurements.\footnote{For earlier work along these lines, see Refs.~\cite{LRV,CFKV,FGKM}.}

The paper is organized as follows. In Section~\ref{sec:CW}, we present the general argument for the correlation between naturalness and loop-induced Higgs couplings to gluons and photons. The key observation is that the same object, the Higgs-dependent mass of the top partner (or partners), determines the dominant radiative corrections to the Higgs mass parameter, via the Coleman-Weinberg (CW) potential, and the top partner contributions to the Higgs couplings to gluons and photons, via the well-known ``low-energy theorems"~\cite{LET}. In Section~\ref{sec:1partner}, we study the correlation between fine-tuning and Higgs couplings quantitatively, using a simple toy model with a single top partner (scalar or fermion) as the benchmark. In Section~\ref{sec:2partners}, we explore how the picture may be affected by the presence of a second top partner, and find that excepting small regions of parameter space where accidental cancellations occur, the conclusions of the benchmark one-partner analysis remain valid. We discuss our findings and conclude in Section~\ref{sec:concl}. 

\section{General Argument: Top Partners, Naturalness, and the Higgs Couplings}
\label{sec:CW}

The starting point of our analysis is a single Higgs doublet $H$ with the SM tree-level potential
\beq
V(H) = -\mu^2 |H|^2 + \lambda |H|^4.
\eeq{SMpot} 
This hypothesis is the simplest interpretation of the LHC discovery consistent with all other experimental data. In particular, there is no evidence in the data of $H$ mixing with other scalar fields, and the constraints on such mixing are now quite stringent. In the SM, the measurements of the Higgs vacuum expectation value (vev) and mass provide precise values for the parameters in the potential:
\beq
\mu = 90~{\rm GeV}, \lambda = 0.13.
\eeq{pars}
How natural are these parameters? To address this question, we need to consider quantum corrections to the potential~\leqn{SMpot}. At the one-loop order, these corrections are conveniently given by the Coleman-Weinberg (CW) formula
\beq
V_{\rm CW}(h) \,=\, \frac{1}{2}\sum_k g_k (-1)^{F_k} \int \frac{d^4\ell}{(2\pi)^4}\,\log\left( \ell^2 + m_k^2(h) \right)\,,
\eeq{Vcw}
where the sum runs over all particles in the model, and $g_k$ and $F_k$ is the multiplicity and fermion number of each particle, respectively. For example, for a gauge-singlet complex scalar, $g=2$ and $F=0$; for a gauge-singlet Dirac fermion, $g=4$ and $F=1$. Here $h/\sqrt{2}$ is the real part of the $U(1)_{\rm em}$-neutral component of $H$; in the SM vacuum, $\langle h \rangle = 
246$ GeV. The one-loop correction to the Higgs mass parameter is given by 
\beq
\delta \mu^2 \equiv \frac{\delta^2 V_{\rm CW}}{\delta h^2}|_{h=0}.
\eeq{dmu}
In the SM, the largest contribution to the CW potential comes from the top quark, since the top Yukawa is the strongest coupling of the Higgs:
\beq
\delta \mu^2 \,=\,-\frac{3y_t^2}{8\pi^2} \Lambda^2 + \ldots,
\eeq{quad}
where $\Lambda$ is the scale at which all loop integrals in $V_{\rm CW}$ are cut off. Since we expect $\Lambda\gg M_{\rm EW}$, the quantum correction to $\mu$ from the top loop is unreasonably large, and would require fine-tuning if no new physics is present.
If the theory is weakly coupled at the TeV scale, the only way to avoid fine-tuning is to introduce a new particle, the {\it top partner}, with mass at or below the TeV scale. (Multiple top partners may be involved in the divergence cancellation.) Such partners can be spin-0 scalars, as in supersymmetric (SUSY) models\footnote{The special role played by the stops, the partners of the top quarks, in determining the degree of naturalness of the electroweak scale in SUSY models was emphasized in Refs.~\cite{NSUSYold}, and more recently in Refs.~\cite{GoldenSUSY,NSUSY}.}, or vector-like spin-1/2 fermions, as in Little Higgs~\cite{LH,LHreviews} or 5-dimensional composite Higgs models~\cite{5DHiggs}.\footnote{In principle, a spin-1 top partner is also a possibility~\cite{spin1}; we will not consider this case here.} In either case, the top partner mass has the form
\beq
m^2(T_i) = m_{0,i}^2 + c_i h^2 + \cdots 
\eeq{TPmass}
where we allow for the possibility of multiple top partners labeled by $T_i$, and drop the terms of higher order in $h$. By dimensional analysis, such higher-order terms need to be suppressed by powers of a mass scale; our approximation is valid if this mass scale is $\gg v$. The absence of a term linear in $h$ in the mass is a consequence of the top partners' vector-like $SU(2)$ charges. The combined top sector contribution to the quadratic terms in the Higgs potential is
\beq
\delta\mu^2 \,=\, \frac{1}{16\pi^2} \left[ \left( \sum_i g_i (-1)^{F_i} c_i - 6 y_t^2 \right)\Lambda^2 + \sum_i g_i (-1)^{F_i} c_i m_{0,i}^2 \log \frac{\Lambda^2}{m_{0,i}^2} - 6y_t^2 m_t^2\log \frac{\Lambda^2}{m_{t}^2}  + \ldots \right]\,.
\eeq{CWtops}
Cancellation of the quadratic divergence yields the sum rule
\beq
6y_t^2 = \sum_i g_i (-1)^{F_i} c_i\,.
\eeq{sumrule}
This sum rule is imposed by the symmetry of the theory in both SUSY and Little Higgs. The remaining fine-tuning can be quantified by taking the ratio of the quantum correction to $\mu^2$ to its measured value:
\beq
\Delta = \frac{\delta\mu^2}{\mu^2} \approx 0.78 \left( \sum_i g_i (-1)^{F_i}c_i \left( \frac{m_{0,i}}{1~{\rm TeV}} \right) ^2 \log \frac{\Lambda^2}{m_{0,i}^2} - 6y_t^2 \left(\frac{m_{t}}{1~{\rm TeV}}\right)^2 \log \frac{\Lambda^2}{m_{t}^2} \right)\,.
\eeq{FTmeasure} 
If $\Delta\gg 1$, the theory must be fine-tuned to accommodate the observed EWSB. Note that $\Delta$ only measures fine-tuning in the Higgs mass parameter; we assume that the observed quartic coupling can be generated with no additional fine-tuning. In certain specific models, such as the minimal supersymmetric standard model (MSSM), a significant loop contribution to the quartic is required, which in turn implies strong constraints 
on the top sector and associated fine-tuning in the Higgs mass parameter. However, such correlations between the quartic and the mass are very model-dependent, and we will not take them into account in this analysis. 

The effects of the top partners on the Higgs couplings first appear at the one-loop level. The best place to look for such effects is in the couplings which vanish in the SM at the tree level. We focus on the couplings of the Higgs to gluons and photons. At the one-loop order, the contributions of particles with masses $\gg m_h$ to these couplings are described by effective operators,
%N: Modified to take out the W contributions (the reasoning being that we want a theory effective at the scale where we integrate out the tops. The W's are still dynamical degrees of freedom at that scale, so their contribution still has to be added "by hand".
\beq
{\cal L}_{h\gamma \gamma} \,=\, \frac{2 \alpha}{9 \pi v} C_\gamma h F_{\mu \nu} F^{\mu \nu}\,,~~~
{\cal L}_{hgg} \,=\, \frac{\alpha_s}{12 \pi v} C_g h G_{\mu \nu} G^{\mu \nu}\,.
\eeq{Hlet}
The Wilson coefficients can be found using the well-known ``low-energy theorems"~\cite{LET}:
\beqa
C_\gamma &=& 1 + \frac{3}{8}\sum_f^{Dirac\; fermions} N_{c,f}Q_f^2 \frac{\partial \ln m_f^2(v)}{\partial \ln v} + \frac{3}{32}\sum_s^{scalars} N_{c,s}Q_s^2 \frac{\partial \ln m_s^2(v)}{\partial \ln v} \,,\CR
C_g &=&  1 + \sum_f^{Dirac\; fermions} C(r_f) \frac{\partial \ln m_f^2(v)}{\partial \ln v} + \frac{1}{4}\sum_s^{scalars} C(r_s) \frac{\partial \ln m_s^2(v)}{\partial \ln v}\,,
\eeqa{Cs}
\\
%N: Modified to remove references to Ws.
where the first term is the contribution of the SM top loops, the sum runs over the top partners, and $N_{c,i}$ and $Q_i$ are the dimension of the $SU(3)_c$ representation and the electric charge (in units of electron charge) of the particle $i$. Note that the exact same objects, the Higgs-dependent masses of top partners $m_i(h)$, enter the CW potential and the Higgs couplings, providing a very general and robust connection between these quantities. In the approximation of Eq.~\leqn{TPmass}, we obtain
%N: Modified to reflect the remove of the W from above.
\beqa
C_\gamma &\approx& 1 + \frac{3}{4}\sum_f  \frac{N_{c,f}Q_f^2 c_fv^2}{m_{0,f}^2 + c_f v^2} + \frac{3}{16}\sum_s  \frac{N_{c,s}Q_s^2 c_s v^2}{m_{0,s}^2 + c_sv^2}\,,\CR
C_g &\approx&  1  +2 \sum_f  \frac{C(r_f) c_fv^2}{m_{0,f}^2 + c_f v^2} + \frac{1}{2}\sum_s \frac{C(r_s) c_s v^2}{m_{0,s}^2 + c_sv^2}\,.
\eeqa{CsTP}
The set of coefficients $\{m_{0,i}, c_i\}$ determines both the fine-tuning $\Delta$ and the Wilson coefficients, generically resulting in a correlation between these quantities. Assuming that there are no other non-SM contributions to the Higgs couplings 
to photons and gluons, the deviations of these couplings from the SM in the presence of top partners are given by
\beq
R_g\equiv \frac{g(hgg)}{g(hgg)|_{\rm SM}}\,=\,C_g,~~~R_\gamma\equiv \frac{g(h\gamma\gamma)}{g(h\gamma\gamma)|_{\rm SM}}\,\approx\,1-0.27 \left(C_\gamma-1\right)\,,
\eeq{couplings}
where the contribution of the $W$ loop has been taken into account in the photon coupling.

It should be noted that in the above discussion, we assumed that the top loop contribution to the Higgs couplings is exactly equal to its value in the SM. In some relevant models of new physics, this assumption is not valid, due to deviations of the top Yukawa from its SM value. Models in which the Higgs is a pseudo-Goldstone boson, such as Little Higgs models, provide an example. In these models, the shift in the top loop contribution to $hgg$ and $h\gamma\gamma$ couplings is of the same order as the top partner loop contributions to these couplings~\cite{TopShift}. The effect of the additional shift is model-dependent. In some cases, a cancellation between the top-Yukawa and top partner loop effects may occur, due to the specific structure of the top mass matrix~\cite{Gilad}. In this case, our analysis would not apply. Note, however, that in all theoretically motivated examples that we are aware of, the shift in the top Yukawa is due to Higgs compositeness, at a scale not far above the electroweak scale. Such models also predict large, tree-level deviations of the Higgs couplings to $W$ and $Z$ bosons, which will be probed with high precision by any experiment capable of precise measurements of gluon and photon couplings. 

\section{Benchmark Model: a Single Top Partner}
\label{sec:1partner}

\begin{figure}[tb]
\begin{center}
\centerline {
\includegraphics[width=3in]{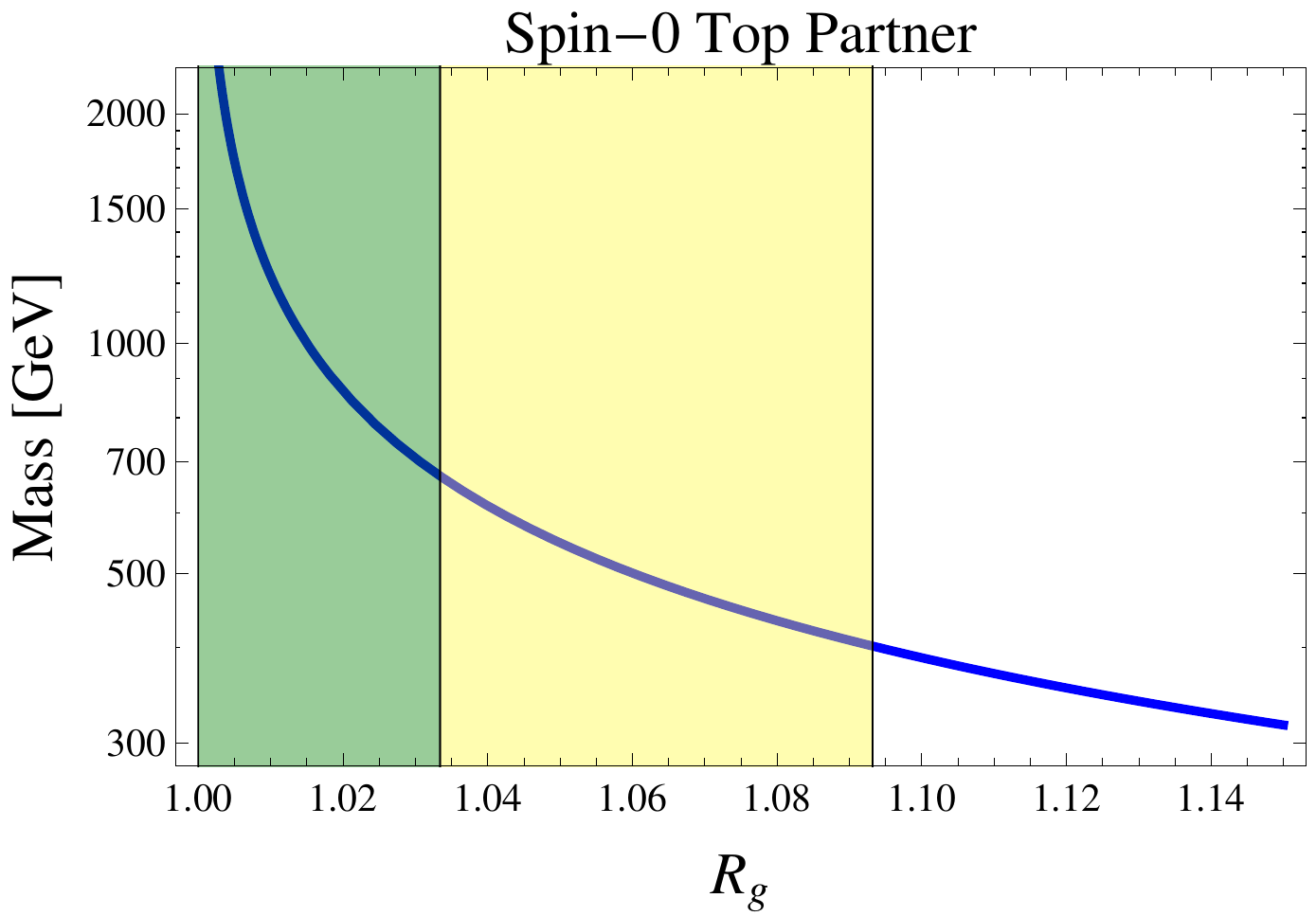} 
\includegraphics[width=3in]{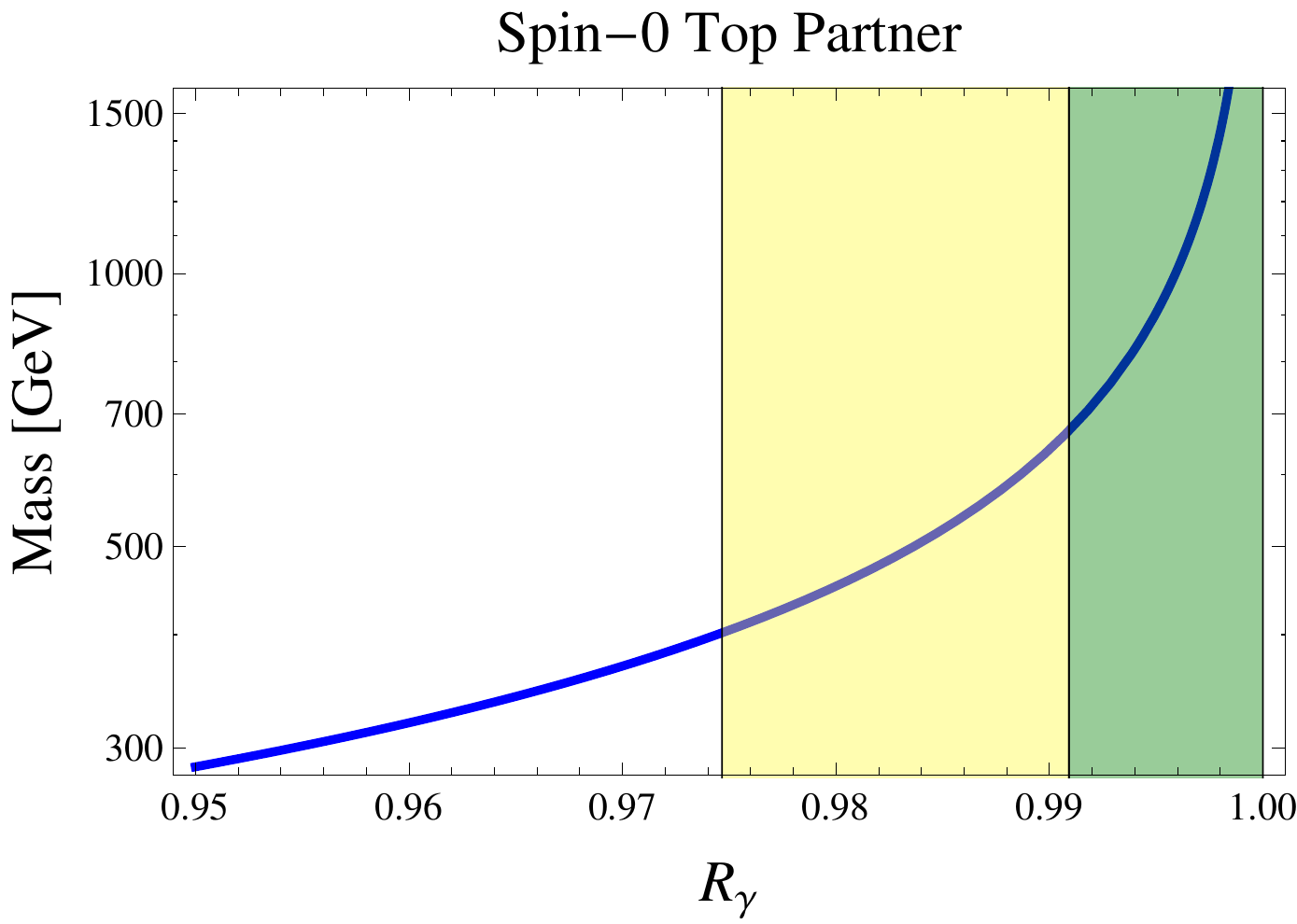}
} 
\centerline {
\includegraphics[width=3in]{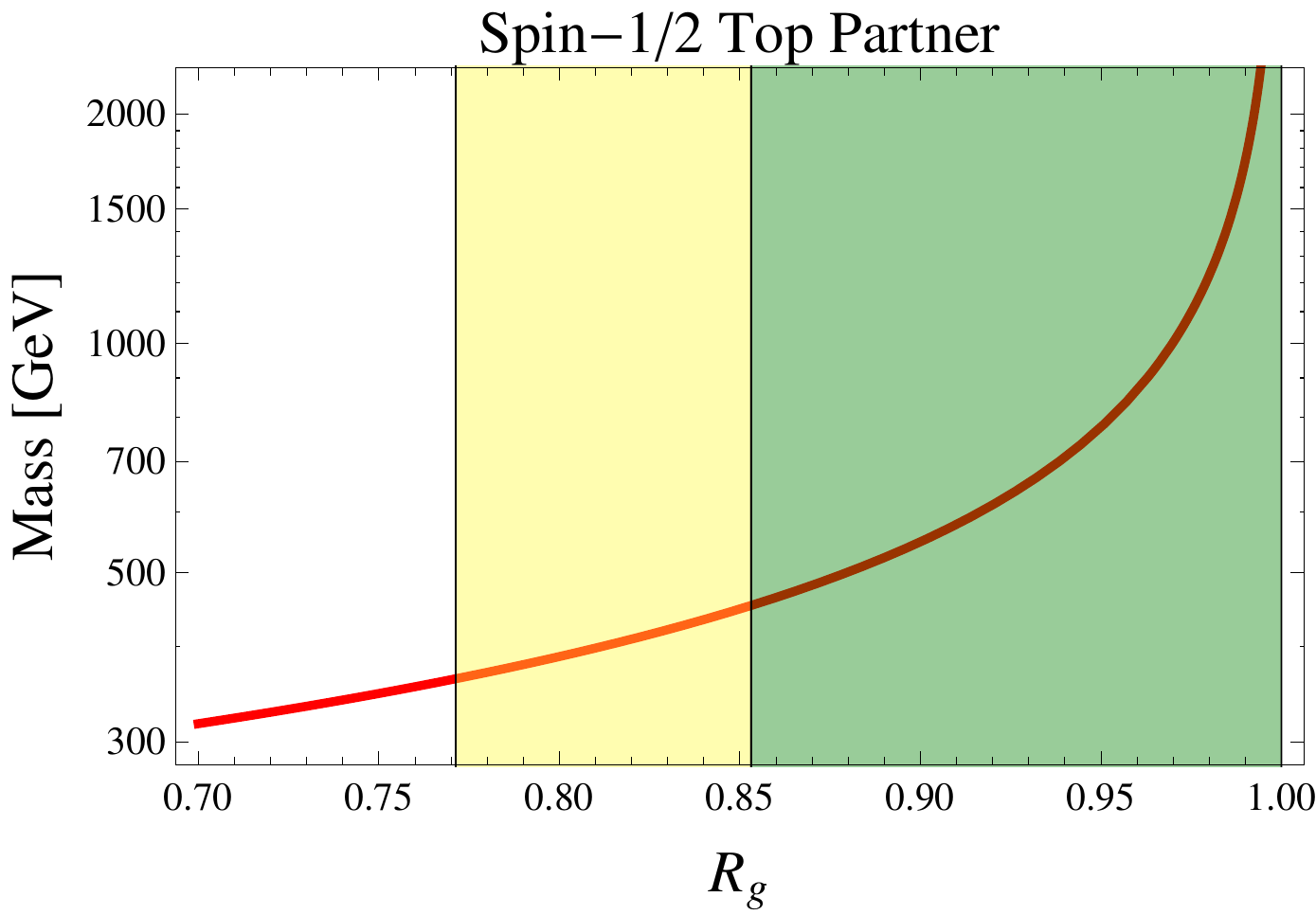} 
\includegraphics[width=3in]{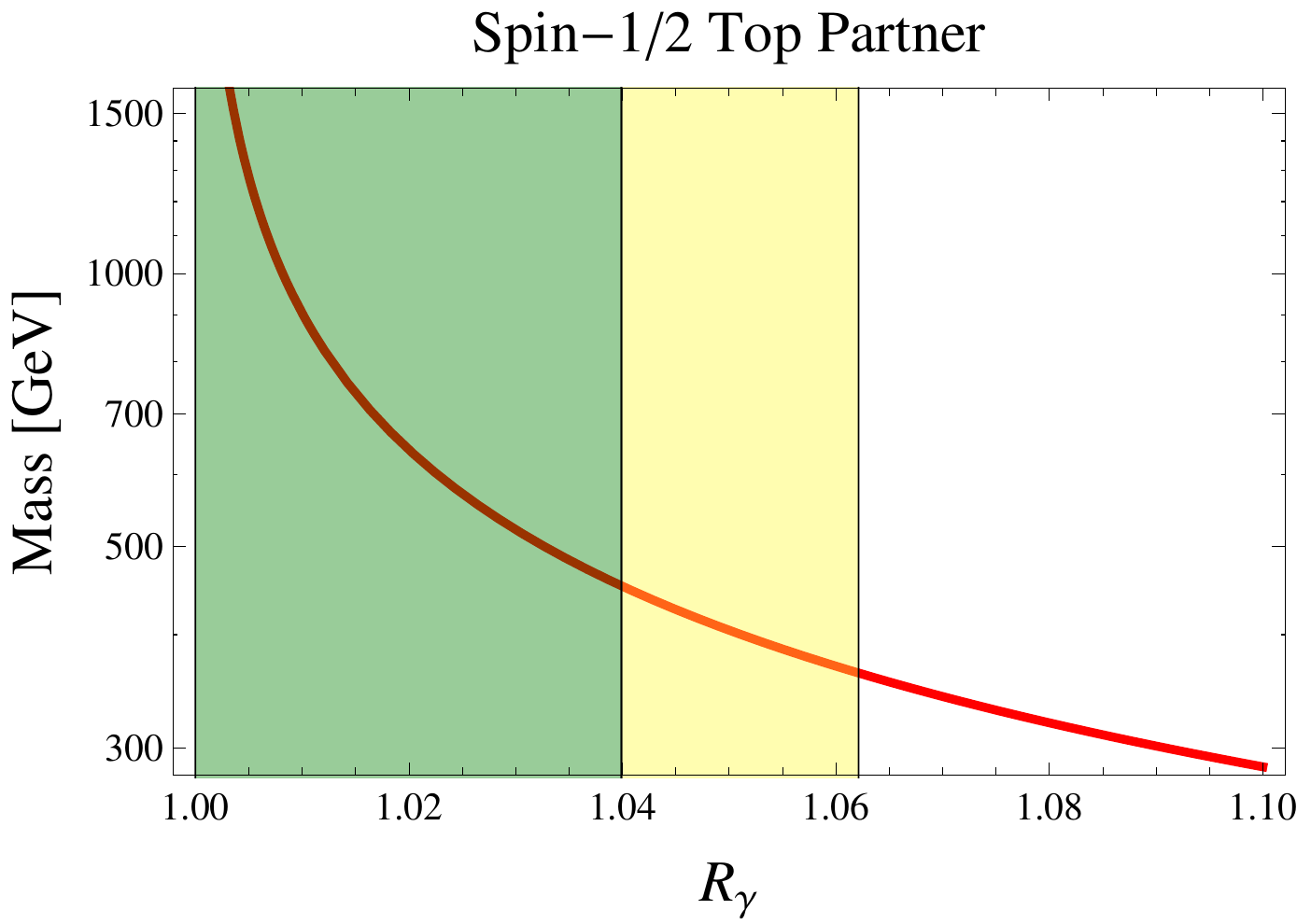}
}
\caption{Fractional deviation of the Higgs coupling to gluons (left panel) and photons (right panel) from the SM value, as a function of the top partner mass. Top row: Spin-0 top partner. Bottom row: Spin-1/2 top partner. Regions currently allowed by the LHC and Tevatron data are shown in green (68~\% c.l.) and yellow (95~\% c.l.).}
\label{fig:Rmass}
\end{center}
\end{figure}

The simplest possibility is that there is a single top partner, in fundamental rep of $SU(3)$ and with electric charge $2/3$, just like the SM top. (The single partner model is applicable to models with multiple top partners if they have the same $m_{0,i}$ parameters: for example, the MSSM with two degenerate stops.) This simple model can be used as a benchmark for evaluating the potential of precision Higgs couplings to probe naturalness. In this case, $c_1$ is fixed by the sum rule~\leqn{sumrule}, and $m_{0,1}$ is the only free parameter in the predictions: 
\beqa
C_\gamma &=& C_g = 1 + \frac{1}{4} \frac{y_t^2v^2}{m_{0,1}^2+y_t^2v^2}~~~~{\rm (spin~0~partner)};\CR
C_\gamma &=& C_g = 1 - \frac{y_t^2v^2}{2m_{0,1}^2-y_t^2v^2}~~~~{\rm (spin~1/2~partner)}.
\eeqa{onepartner}
The correlation between the Higgs coupling deviations from the SM and the mass of the top partner is shown in Fig.~\ref{fig:Rmass}.
For reference, we also show constraints obtained from a fit to the current LHC-7, LHC-8~\cite{Higgs_atlas,Higgs_cms} and Tevatron~\cite{Higgs_tev} data, assuming that top-partner loops are the only non-SM contribution to Higgs couplings. To obtain these constraints, we fit to the published Higgs event rates observed in various channels, assuming no correlation between any of the data points, 
and include the theoretical uncertainties provided by the Higgs cross section working group~\cite{Dittmaier:2011ti}. Our results are roughly consistent with the more detailed fits performed by the LHC collaborations~\cite{ATLAS:2013sla,Chatrchyan:2013lba}: for example, our one-sigma error bar on $R_g$ is about $\pm 0.1$, compared to $0.14$ reported by the ATLAS collaboration~\cite{ATLAS:2013sla} in a two-parameter fit where $R_g$ and $R_\gamma$ were assumed to be the only non-SM contributions to the Higgs rates. A broad range of top partner masses in the region motivated by naturalness are currently allowed by data: the 95\%~c.l. limit on the top partner mass is about 320 GeV for a spin-0 top partner, and 400 GeV for a spin-1/2 partner. (Note that our best-fit value for $R_g$ is about $0.7\sigma$ below the SM expectation of 1.0, resulting in a slightly stronger bound on the spin-0 partners and a slightly weaker bound on the spin-1/2 case.) However, future precise measurements of the Higgs coupling at the LHC-14 and a future $e^+e^-$ facility would probe much of the interesting parameter space. For example, a 1\% measurement of the gluon coupling will probe the top partner masses in excess of 1 TeV, for both spin-0 and spin-1/2 top partners. 

\begin{figure}[tb]
\begin{center}
\centerline {
\includegraphics[width=3in]{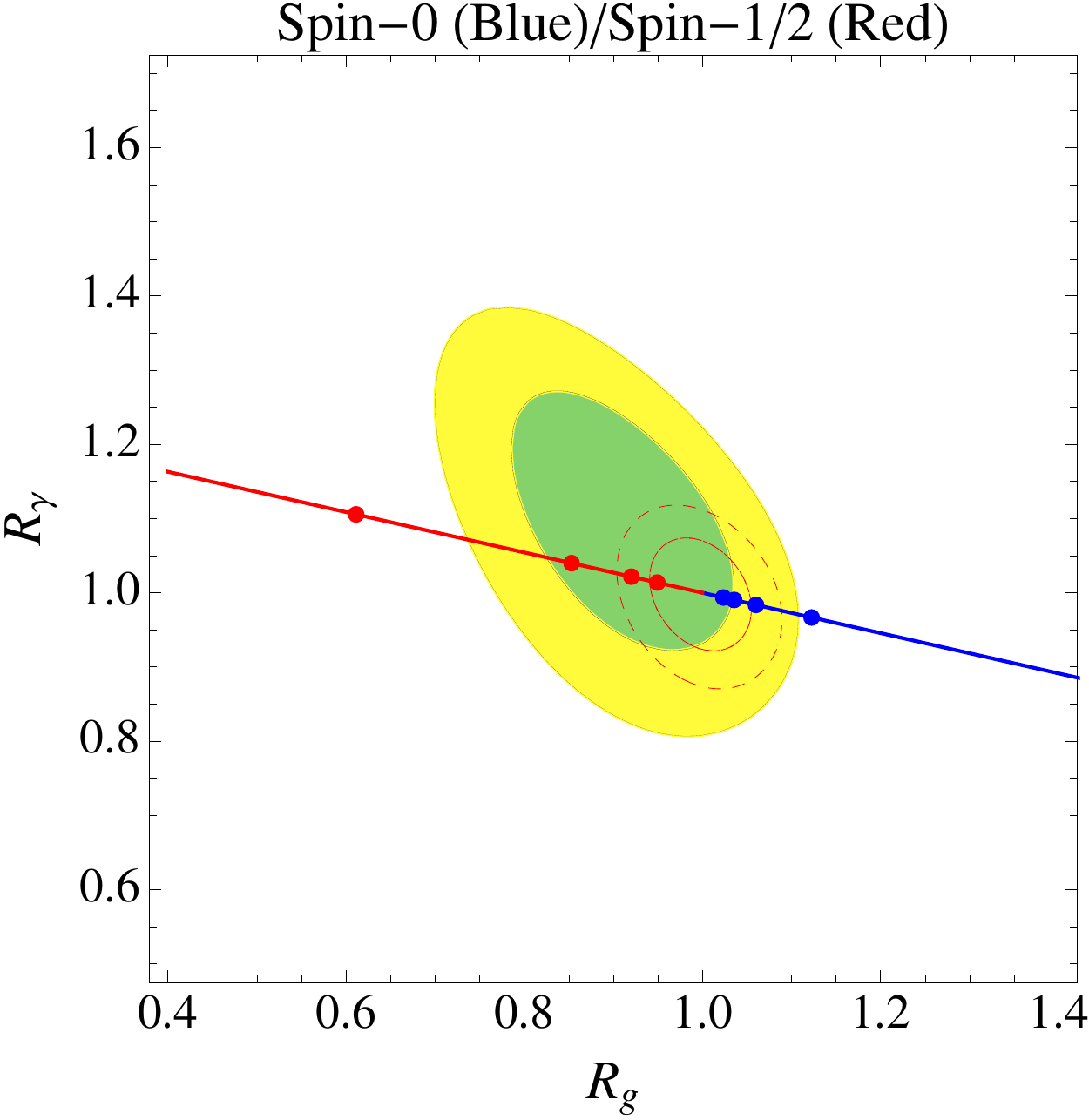}
}
\caption{Regions allowed by the LHC and Tevatron measurements of the Higgs rates in the $R_\gamma-R_g$ plane, at the 68~\% c.l. (green) and 95~\% c.l. (yellow). The spin-0 top partner model predicts deviations along the blue line, while the spin-1/2 top partner induces deviations along the red line. The points on both lines correspond to the partner masses of 350, 500, 650, and 800 GeV. For comparison, projected constraints from the LHC-14~\cite{Peskin} are shown by red lines.}
\label{fig:Traj}
\end{center}
\end{figure}

Since the one-partner model has only one free parameter, the deviations in gluon and photon couplings are correlated. This is shown in Fig.~\ref{fig:Traj}, along with the current and future LHC constraints on the two couplings. (We used the information provided in Ref.~\cite{Peskin} to estimate the LHC-14 contours.) It is clear that the constraints are strongly dominated by the gluon coupling measurement, due to both the slope of the trajectory and the stronger experimental bound on $R_g$. If a deviation from the SM is observed, it would be straightforward to check whether it can be interpreted within the one-partner framework by simply checking whether the trajectories shown here intersect with the experimentally determined region. If the answer is positive, these measurements will also allow to unambiguously determine the top partner spin. 

\begin{figure}[tb]
\begin{center}
\centerline {
\includegraphics[width=3in]{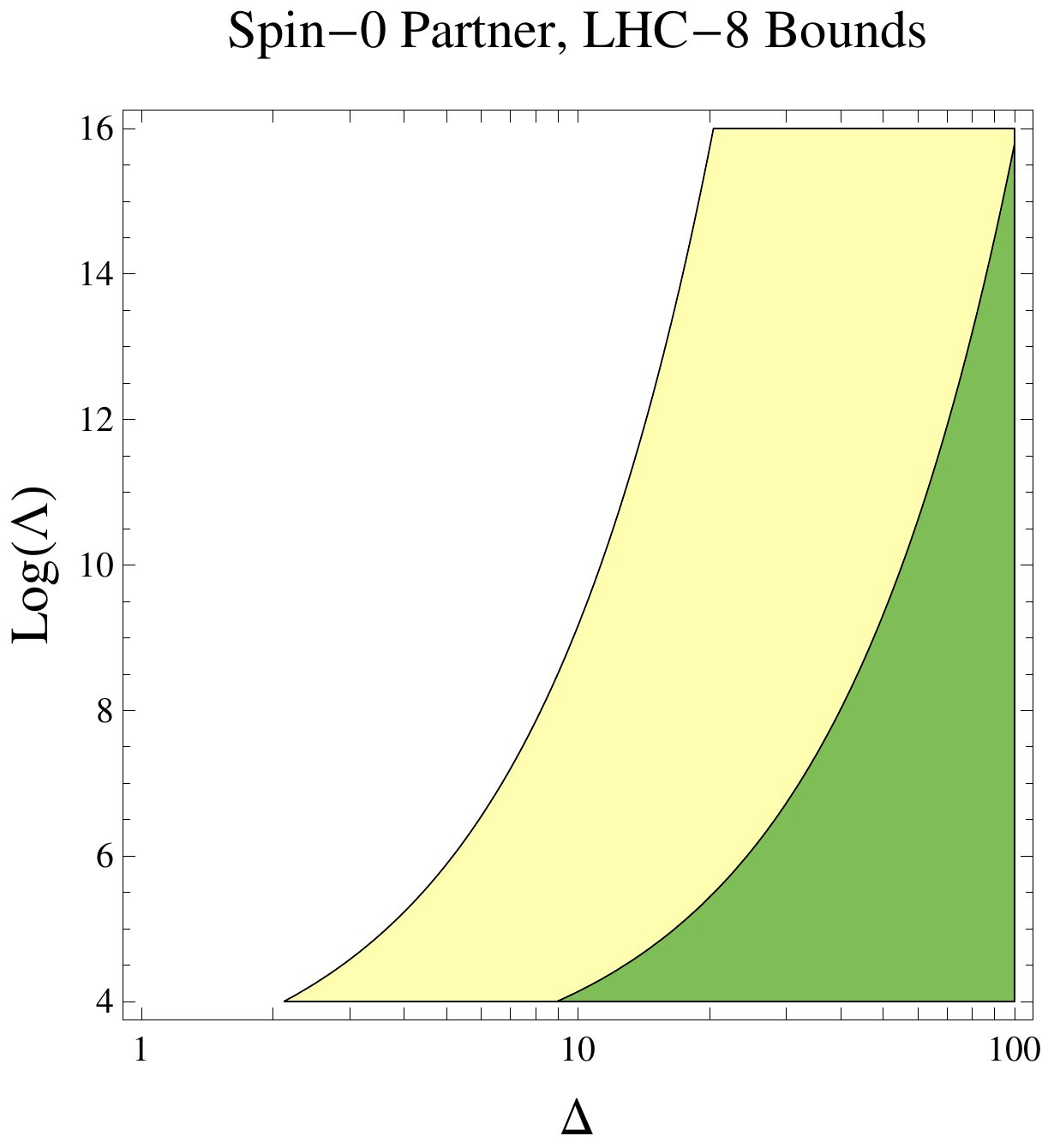}
\includegraphics[width=3in]{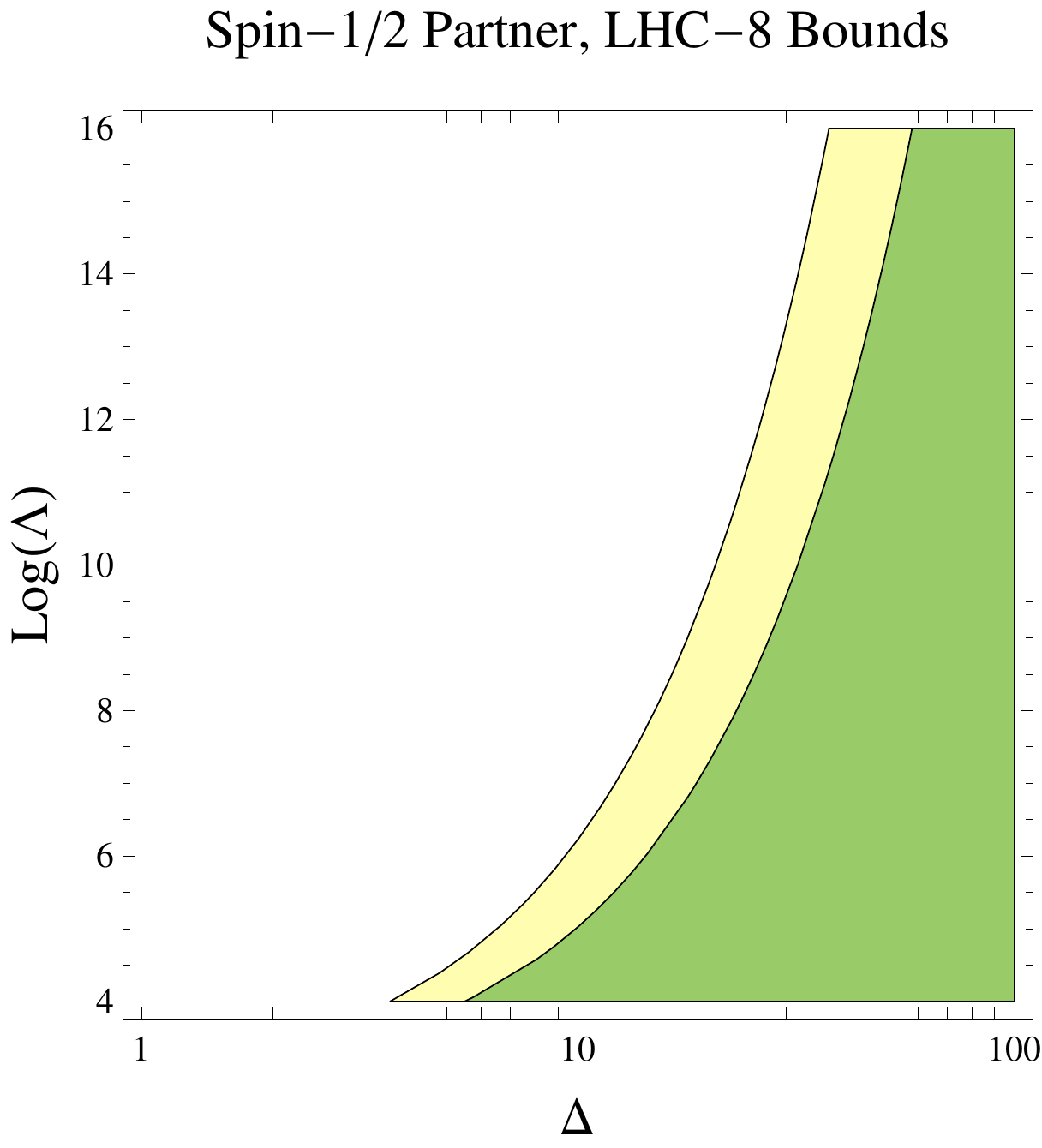}
}
\caption{Regions allowed by the LHC and Tevatron data in the $\Delta-\log\Lambda$ plane, at the 68~\% c.l. (green) and 95~\% c.l. (yellow). Here, $\Lambda$ is the scale (in GeV) where the logarithmic divergence in the Higgs mass renormalization is cut off. Left panel: Spin-0 top partner. Right panel: Spin-1/2 top partner.}
\label{fig:FTtoday}
\end{center}
\end{figure}

\begin{figure}[h!]
\begin{center}
\centerline {
\includegraphics[width=3in]{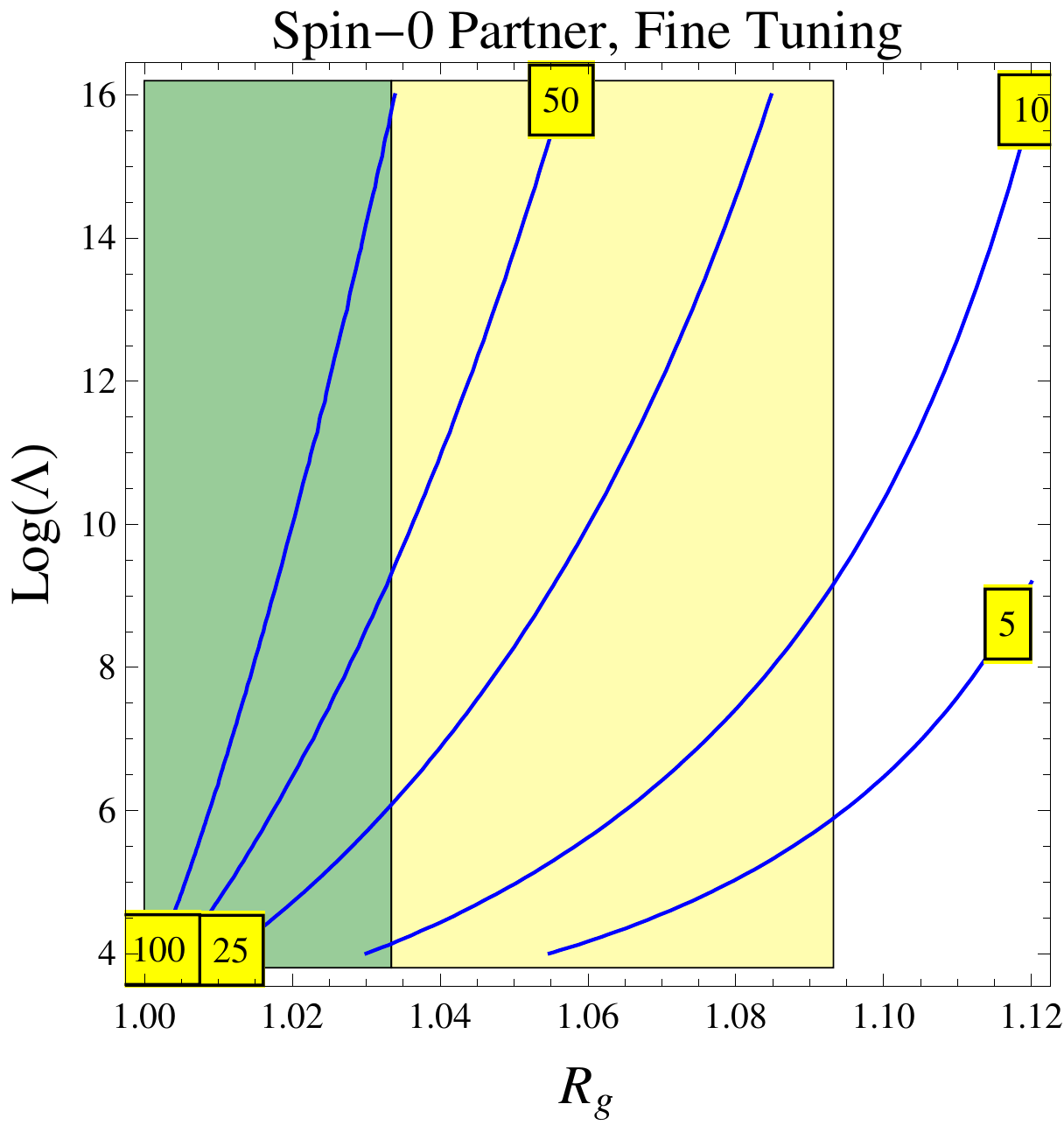}
\includegraphics[width=3in]{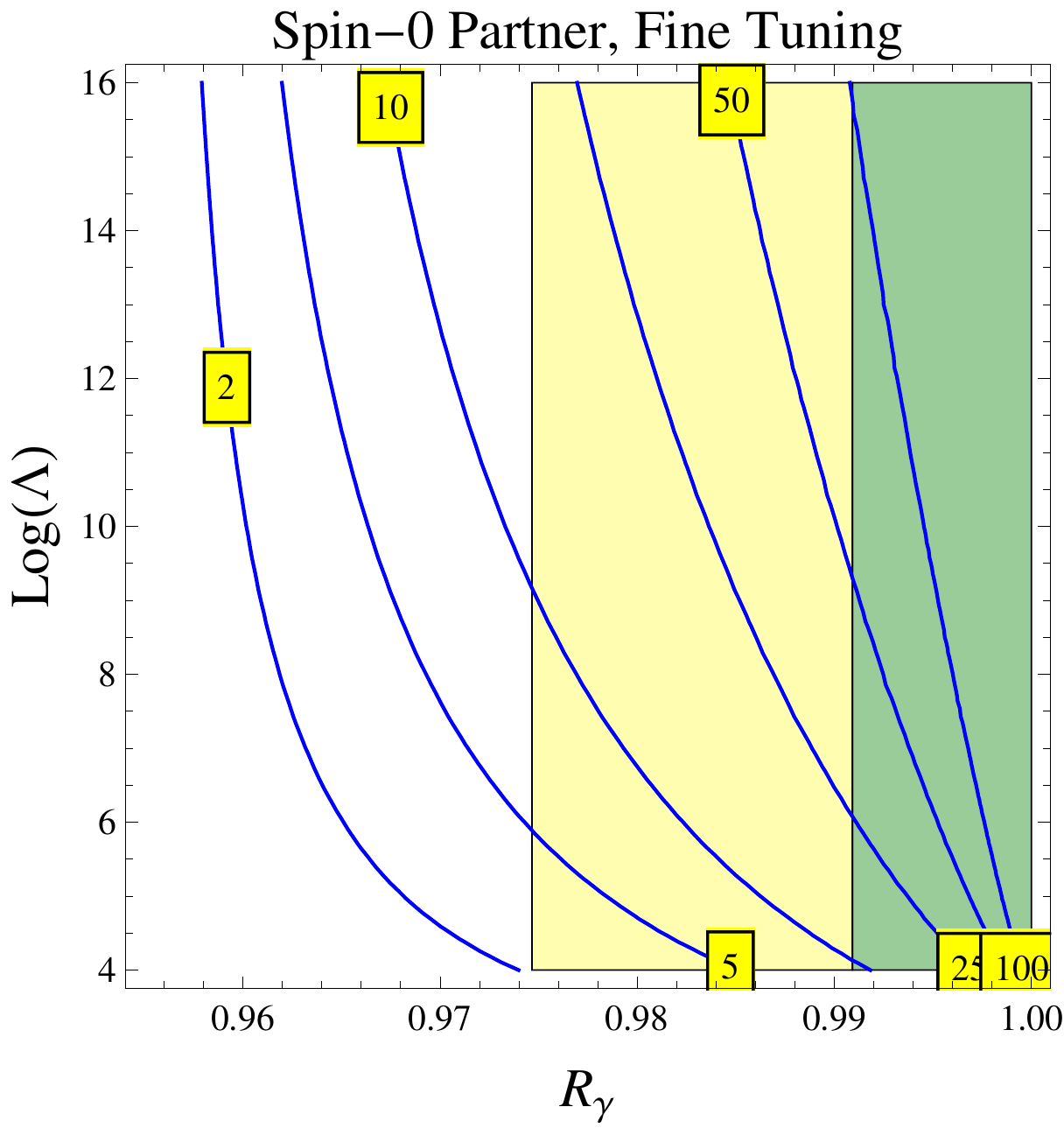}
} 
\centerline {
\includegraphics[width=3in]{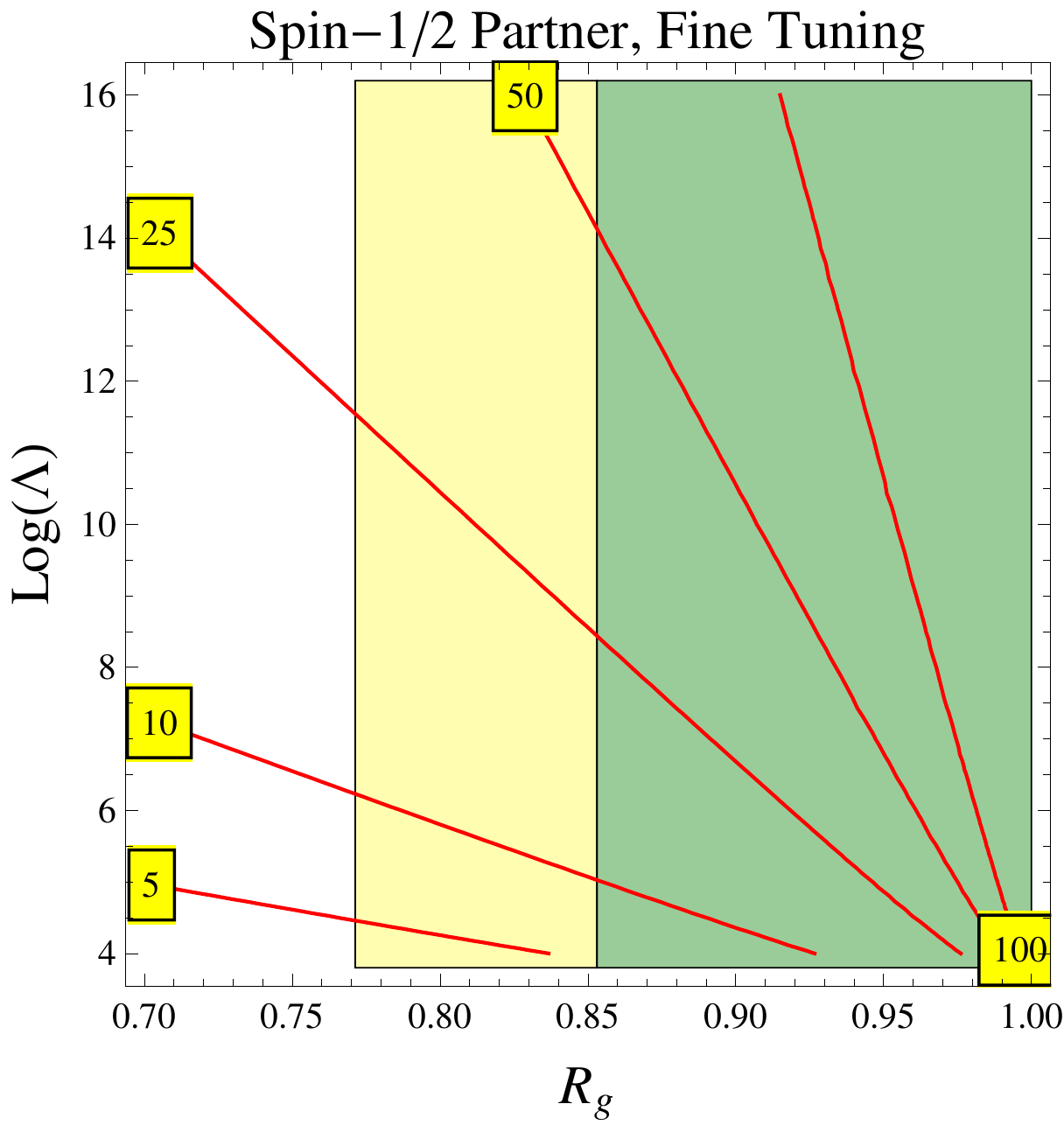}
\includegraphics[width=3in]{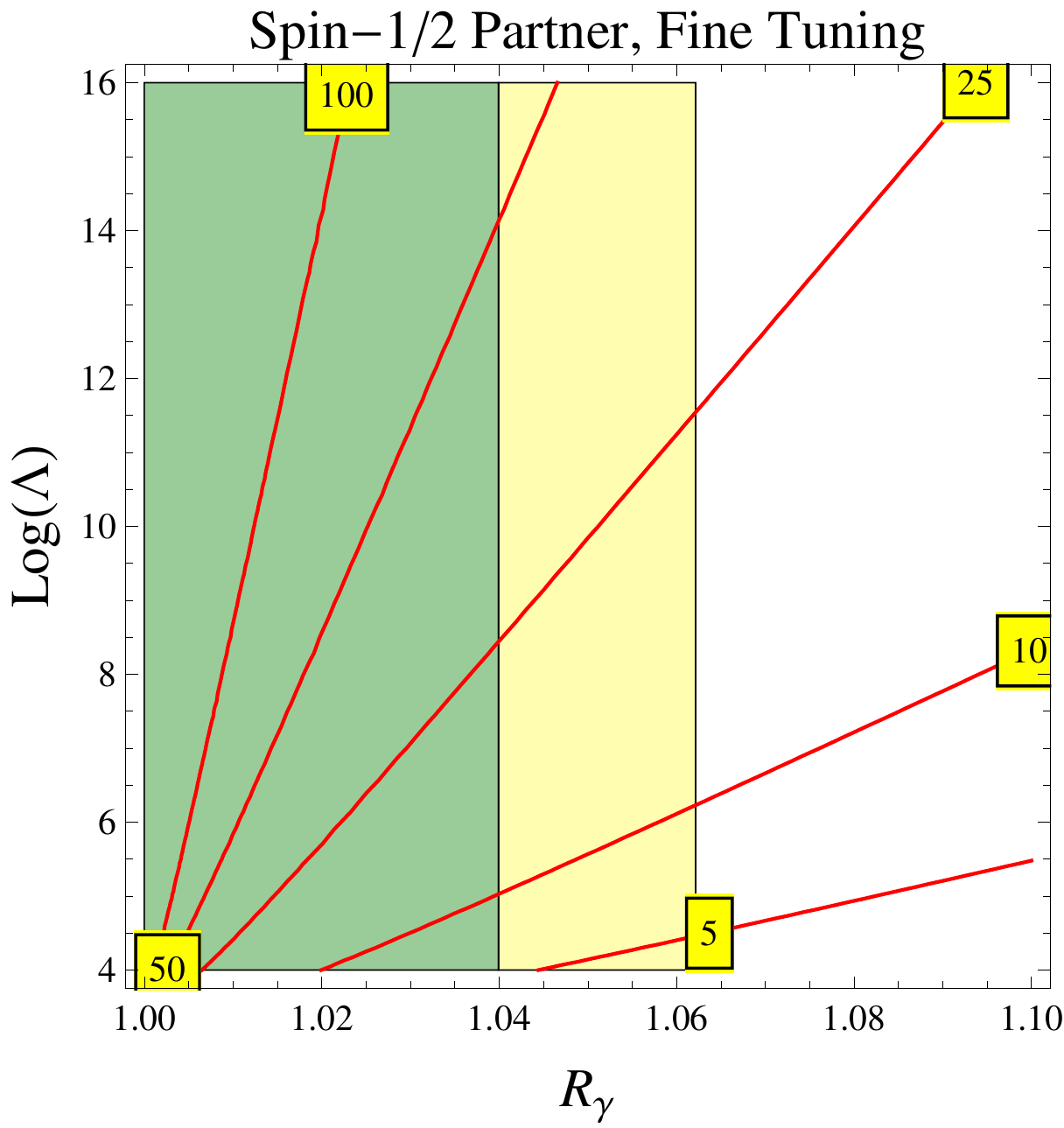}
}
\caption{Fine-tuning as a function of the fractional deviation of the Higgs coupling to gluons (left panel) and photons (right panel) from the SM value, and the energy scale $\Lambda$ (in GeV) where the logarithmic divergence in the Higgs mass renormalization is cut off. Top row: Spin-0 top partner. Bottom row: Spin-1/2 top partner. Regions currently allowed by the LHC and Tevatron data are shown in green (68~\% c.l.) and yellow (95~\% c.l.).}
\label{fig:Rft}
\end{center}
\end{figure}

The connection between Higgs couplings and fine-tuning is illustrated more directly in Figs.~\ref{fig:FTtoday} and~\ref{fig:Rft}. Since the top partners only cut off the quadratic divergence in the top loop, leaving the logarithmic divergence uncanceled, the value of the fine-tuning measure $\Delta$ depends logarithmically on the scale $\Lambda$ where the logarithmic divergence is cut off. The value of $\Lambda$ is very model-dependent. To demonstrate its effect, we vary $\Lambda$ between the ``low" $10$ TeV scale, representing a rough lower bound on this scale in realistic models, and the ``high" $10^{16}$ GeV, motivated by grand unification. 
In the case of a spin-0 partner, the 95\%~c.l. lower bound on the fine-tuning from the current Higgs data varies between $\sim 1/2$ for a low-scale model and $\sim 1/20$ for a high-scale model. The bounds for the spin-1/2 partner are slightly stronger, between $\sim 1/3$ and $\sim 1/30$. Of course, these bounds can be dramatically improved by the future precise measurements of Higgs couplings. For example, if the gluon coupling is found to agree with the SM at a 1\% level, the minimal amount of fine-tuning required would be $\sim 1/25$ for the low-scale model, and $\sim 1/400$ for a high-scale model. 

We emphasize that the probe of naturalness advocated here is complementary to direct searches for top partners. Sensitivity of the direct searches, especially at hadron colliders such as the LHC, depends on details of the spectrum and the decay patterns of the produced top partners. For example, while the LHC ``headline" direct bounds on stops are already about $600-700$ 
GeV~\cite{stopbounds}, these bounds can be evaded in a variety of ways, {\it e.g.} ``stealthy"~\cite{stealthy} or compressed stop spectra, or R-parity violation. In contrast, the nature of the Higgs coupling deviations discussed here is very tightly connected to the restoration of naturalness, and the connection is essentially model-independent. Of course, the simple correlation exhibited in the benchmark one-partner models may be violated in more complicated setups, where for example cancellation among various loop contributions is in principle possible. We will investigate an example of this in the next section. Still, it is worth emphasizing that the ``loopholes" inherent in the test of naturalness proposed here are completely different from the ones plaguing direct searches. Together, these techniques should provide an extremely powerful and robust test of naturalness.

\section{Two Top Partners}
\label{sec:2partners}

\begin{figure}[h!]
\begin{center}
\centerline {
\includegraphics[width=3in]{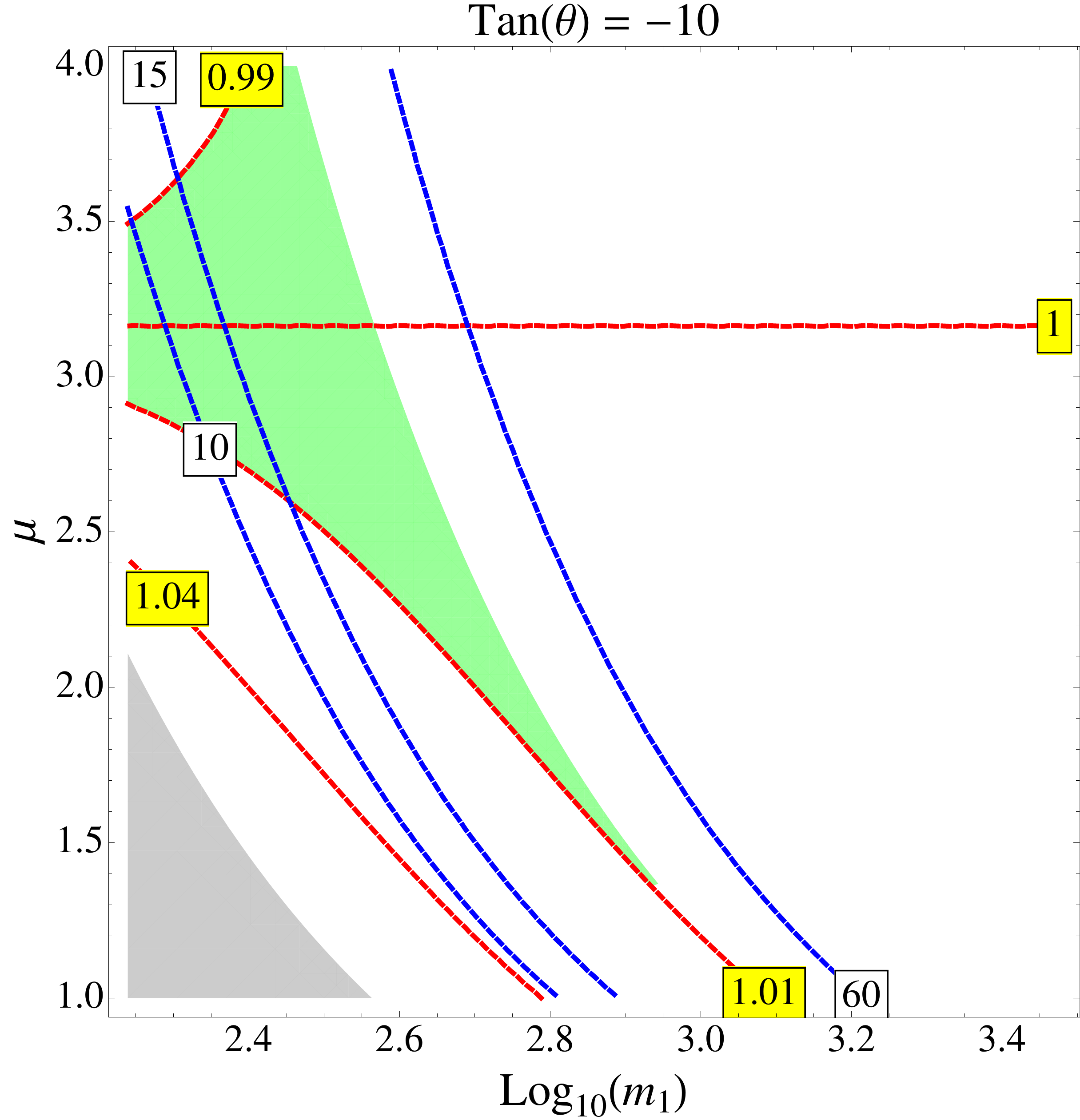} 
\includegraphics[width=3in]{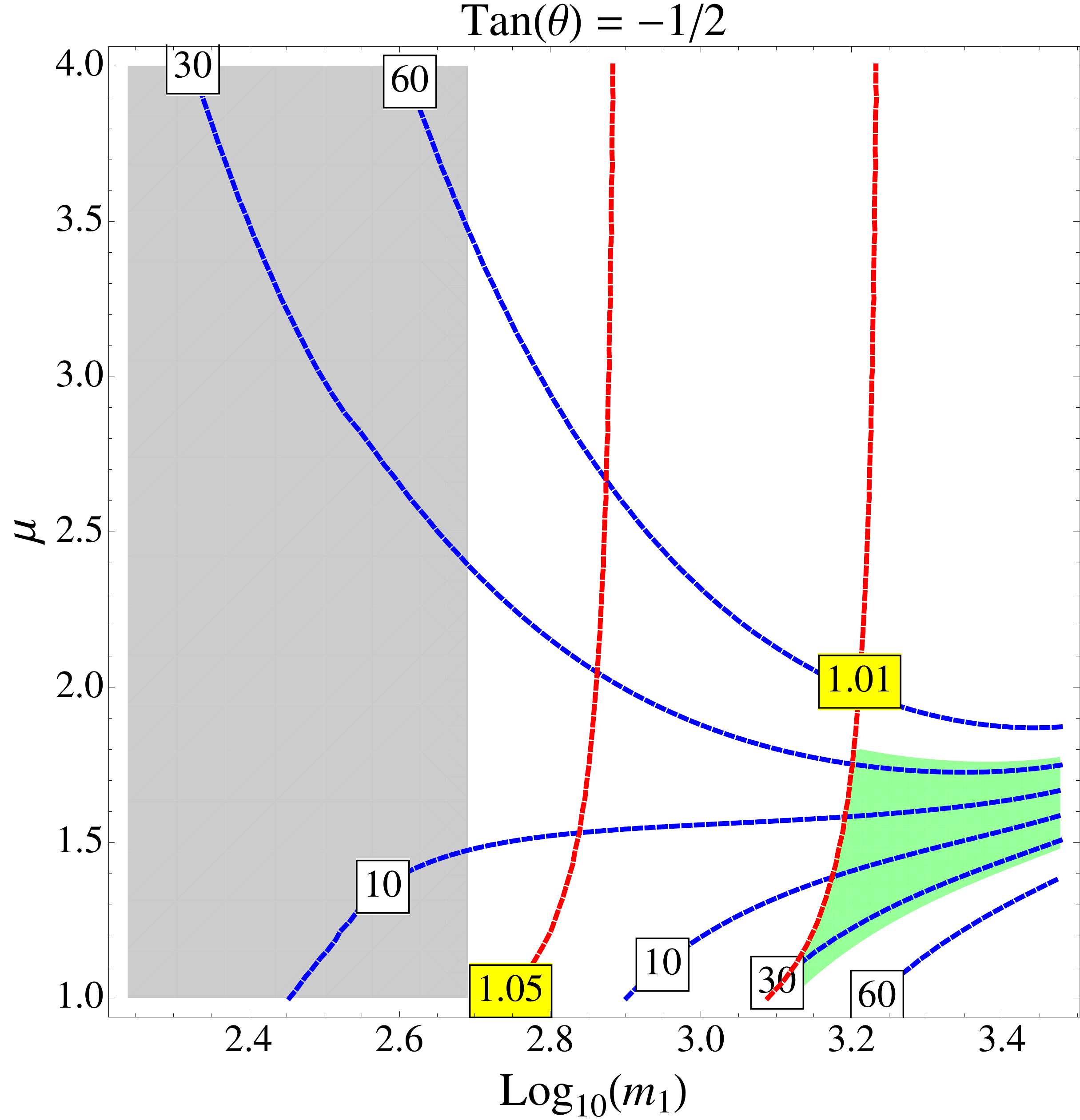}
} 
%\centerline {
%\includegraphics[width=3in]{tantheta0.pdf} 
%} 
\centerline {
\includegraphics[width=3in]{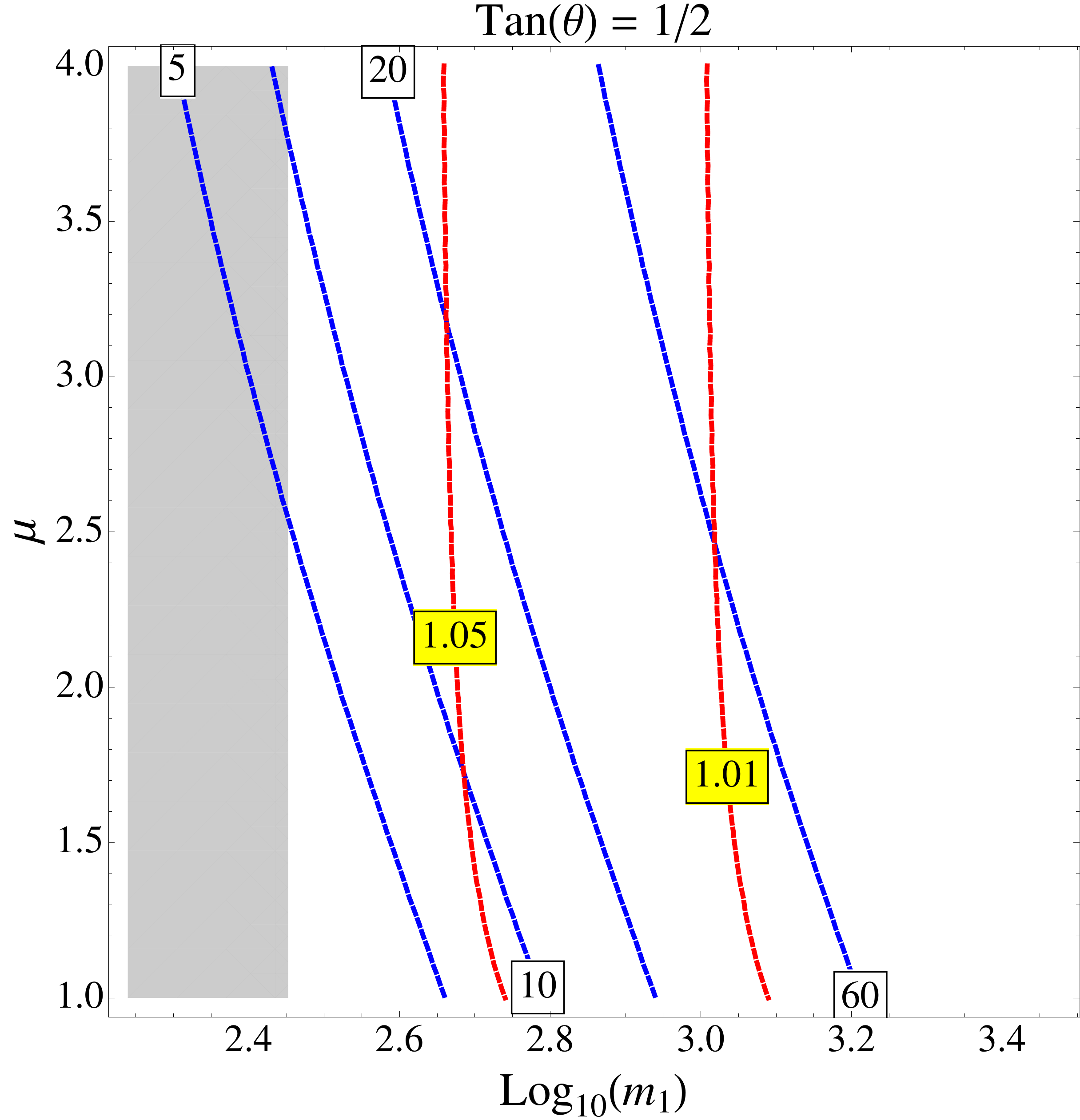} 
\includegraphics[width=3in]{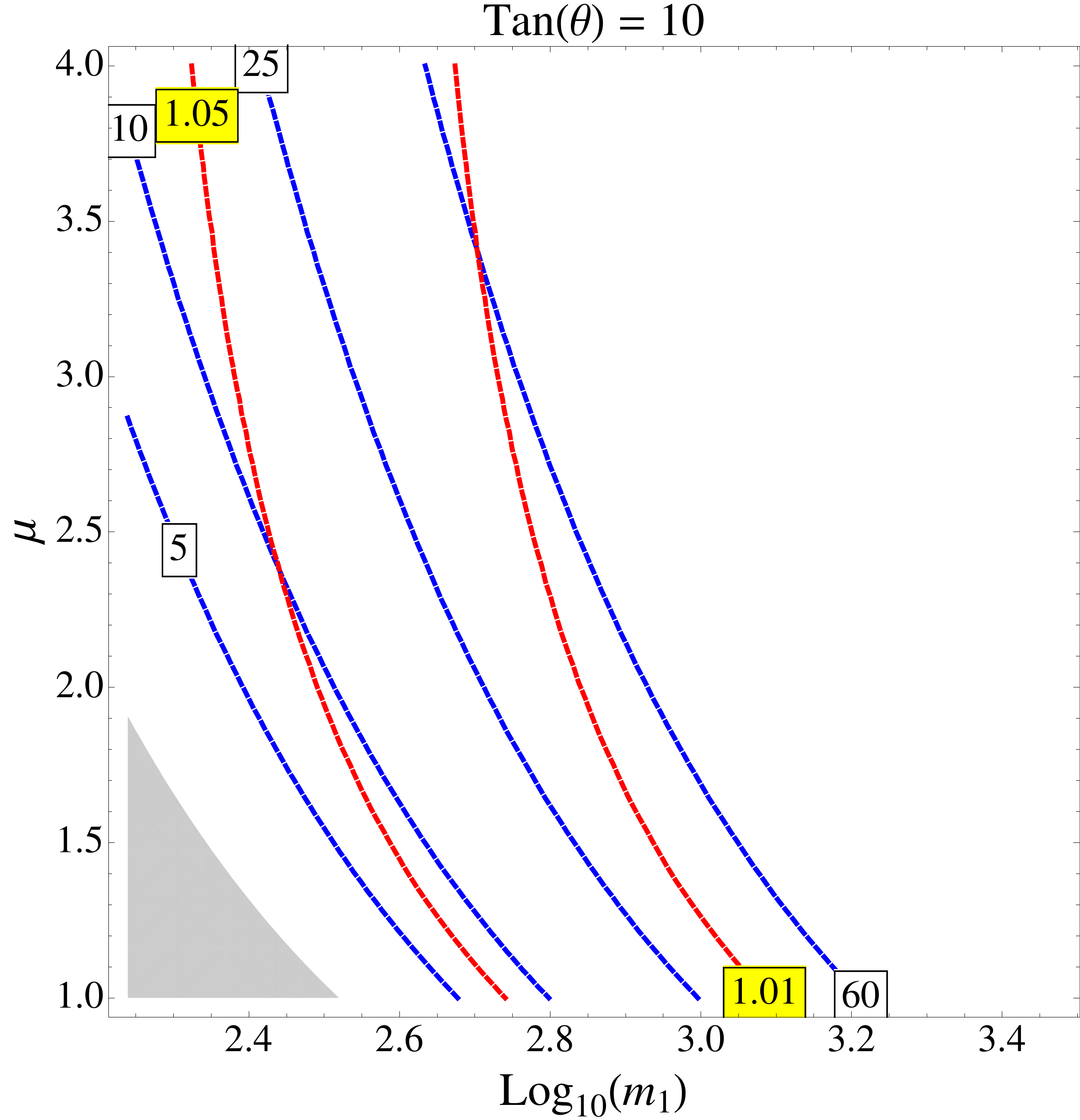}
}
\caption{Contours of fine tuning (in blue) and $R_g$ (in red) for fixed values of $\theta$, with $\Lambda = 20$ TeV. The regions shaded in green correspond to points where $|R_g -1| < 0.01$ but for which the amount of fine tuning is less than what is predicted for a one scalar partner model with $R_g=0.01$. The regions shaded in gray corresponds to points where $|c_i v^2/m_{0,i}^2| >1$. The top partner mass $m_1$ is in units of GeV.}
\label{fig:FTRg2}
\end{center}
\end{figure}

Cancellation of the top loop divergence does not have to be achieved with a single new particle. For example, in the MSSM, there are two spin-0 top partners, $\tilde{t}_1$ and $\tilde{t}_2$, generically with different masses, both of which participate in divergence cancellation. Models with multiple top partners are characterized by multi-dimensional parameter spaces, even after the divergence cancellation sum rule is imposed. We expect that throughout most of the parameter space of a given model, the correlation between Higgs couplings and fine-tuning studied in Section~\ref{sec:1partner} continues to hold. However, there could be special regions of parameter space where it can fail, due to cancellations between contributions of the two top partners, either to the CW potential or to the Higgs couplings. To illustrate this, in this section we will consider a toy model with two spin-0 partners,
both in fundamental rep of $SU(3)$ and with electric charge $2/3$. (These are the quantum number assignments of the MSSM stops, so the results of this section will approximately apply in that model; the correspondence becomes exact in the limit of soft masses large compared to $v$.\footnote{Many authors examined the stop loop contributions to Higgs couplings in the MSSM; see, for example, Refs.~\cite{SUSY}.}) The model has four free parameters, $\{m_{0,i}, c_i\}$, $i=1, 2$; after the sum rule~\leqn{sumrule} is imposed, the number is reduced to 3. We choose to work in terms of
\beq
m_1;~~\mu = \frac{m_2}{m_1};~~\theta = \tan^{-1} \frac{c_2}{c_1}\,,
\eeq{3pars}
where $m_i=\sqrt{m_{0,i}^2+c_i v^2}$ are the physical masses of the top partners, and $m_1<m_2$. Note that in the limits ($\mu\to 1$, any $\theta$) and ($\theta\to 0$, any $\mu$), the model reduces to the one-partner model with the same $m_{0,1}$, considered in Section~\ref{sec:1partner} above.

\begin{figure}[t!]
\begin{center}
\centerline {
\includegraphics[width=3.5in]{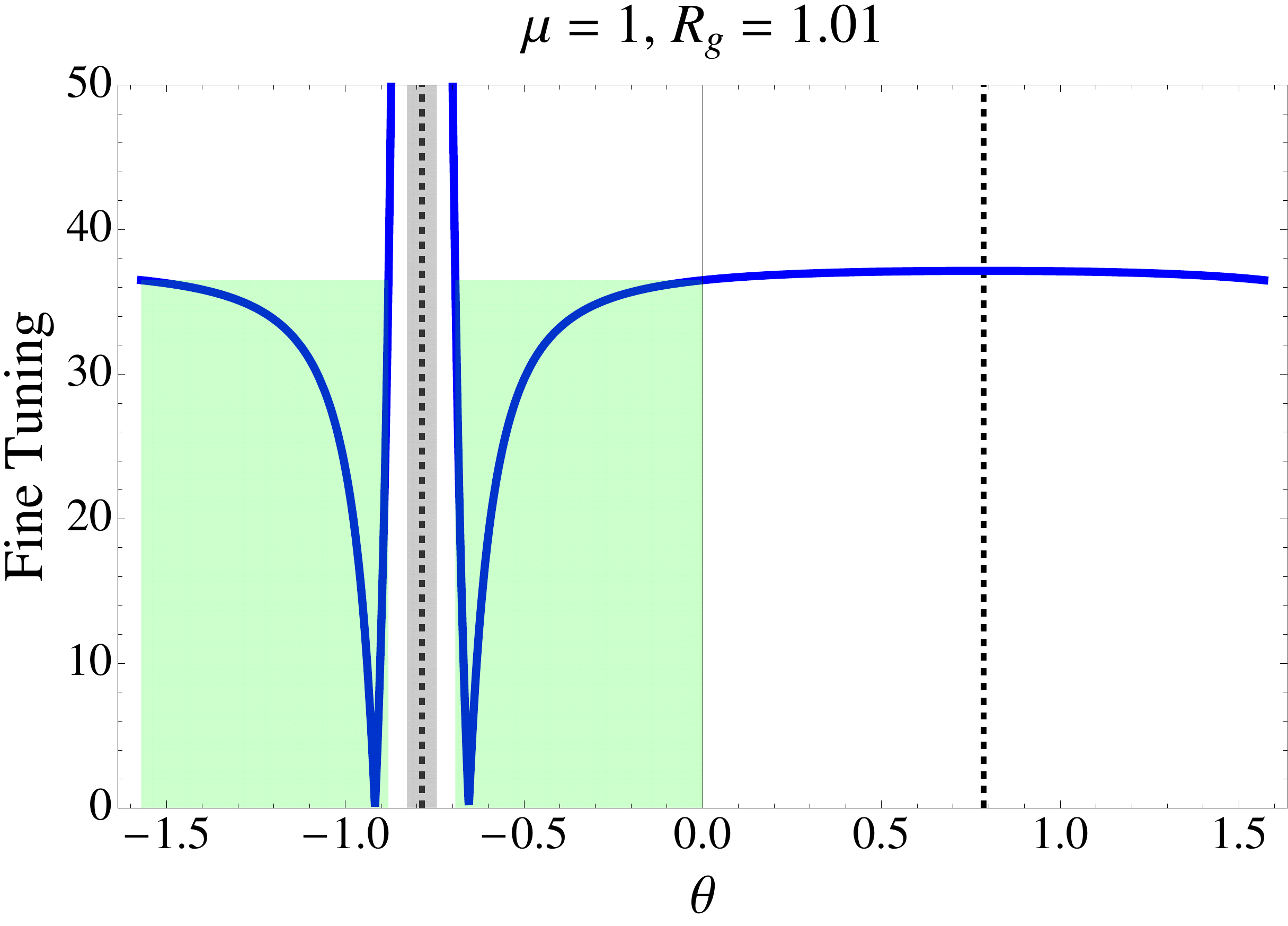} 
\includegraphics[width=3.5in]{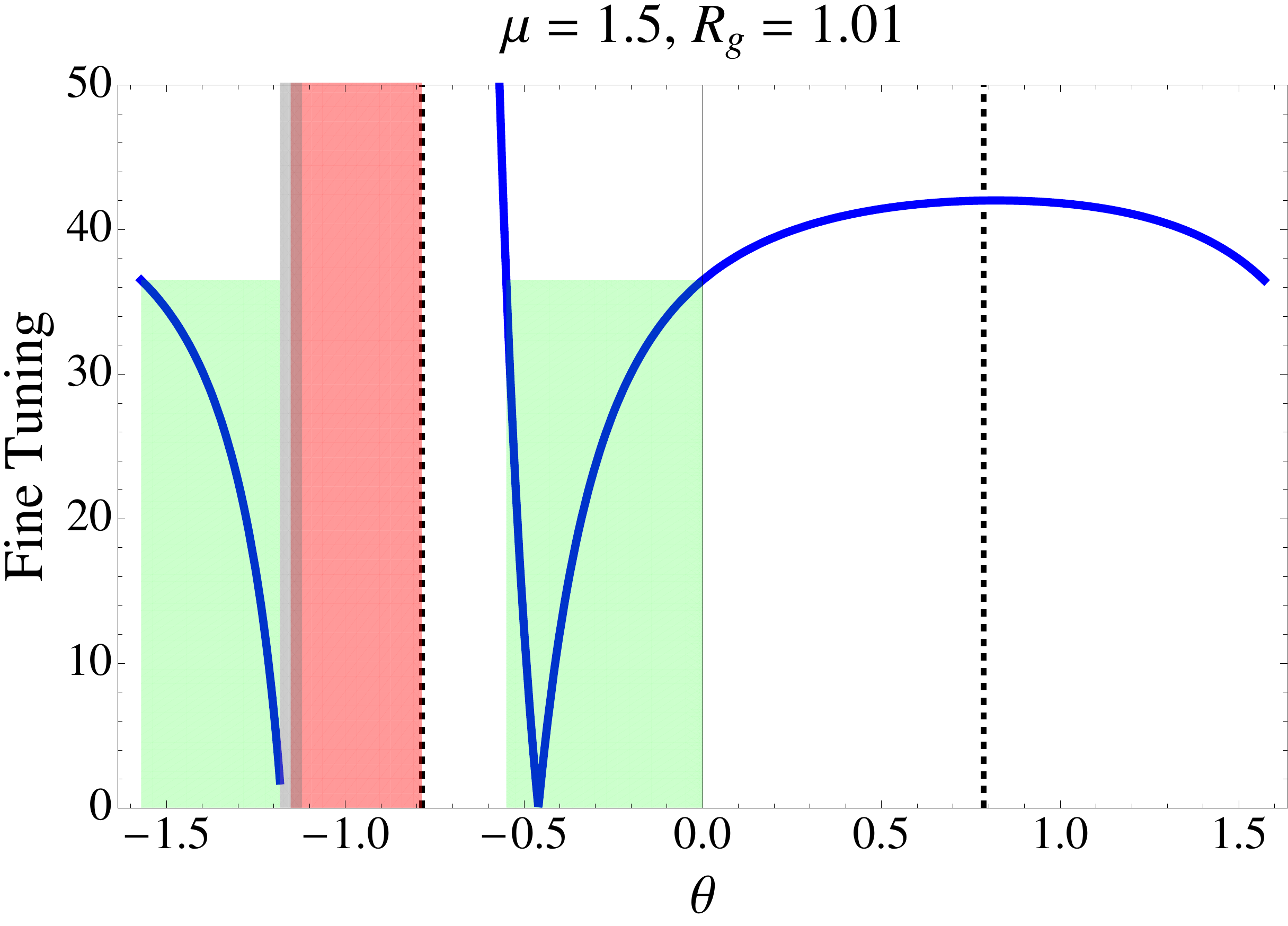}
} 
\centerline {
\includegraphics[width=3.5in]{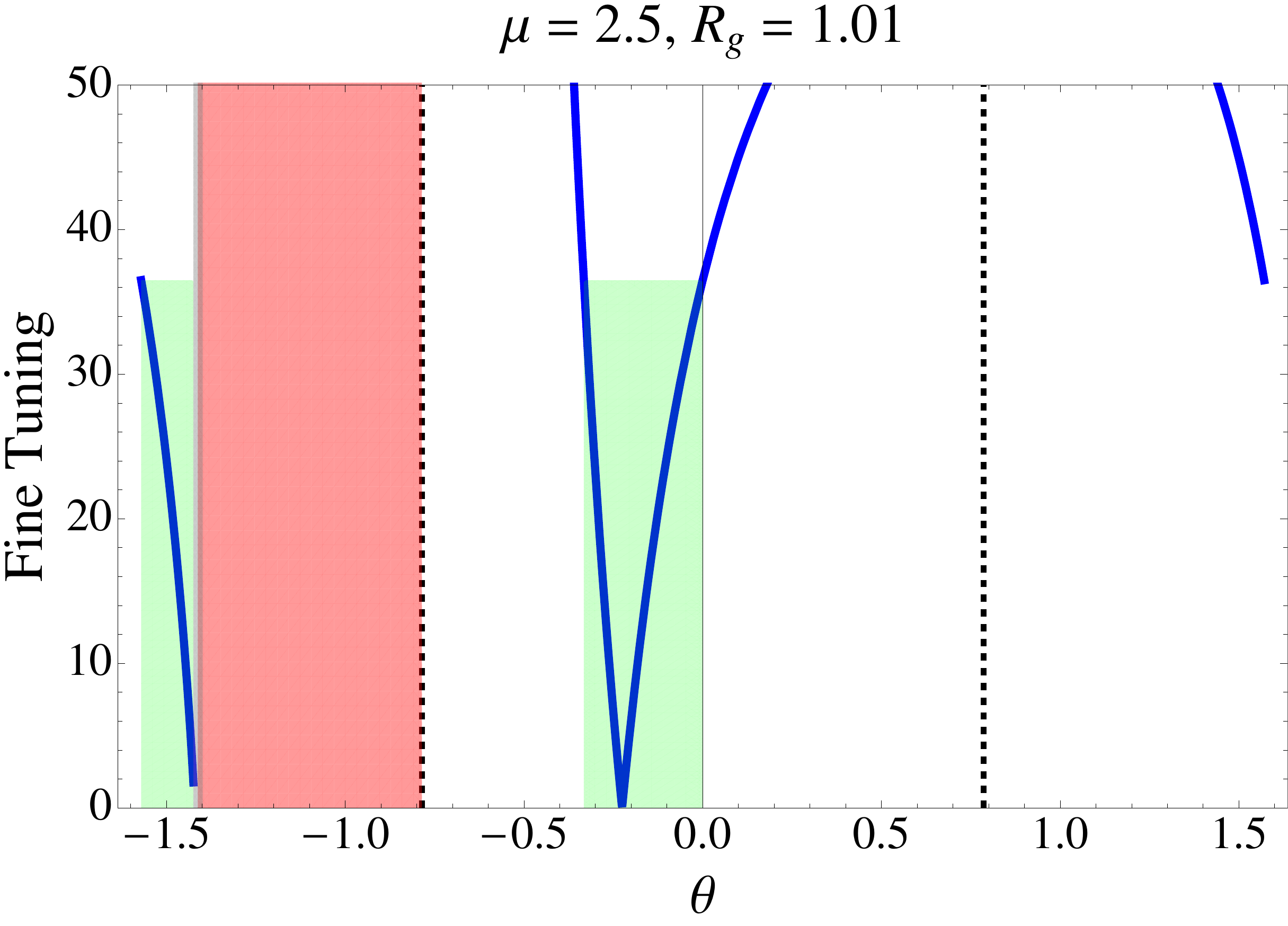} 
} 

\caption{Fine tuning (in blue) as a function of $\theta$ for fixed values of $\mu$ and $R_g$, with $\Lambda = 20$ TeV. The regions shaded in in gray indicate values of $\theta$ where $|c_i v^2/m_{0,i}^2| >1$; the regions shaded in red are unphysical due to $m_1^2 <0$. Green regions indicate values of $\theta$ for which the fine tuning is less than what is predicted for a one scalar partner model with $R_g=0.01$.}
\label{fig:FTtheta}
\end{center}
\end{figure}

The main conclusion of our analysis of this model is that the correlation of the Higgs coupling deviations and fine-tuning, observed in the benchmark one-partner models of Section~\ref{sec:1partner}, is rather robust. This is illustrated in Fig.~\ref{fig:FTRg2}. For example, suppose that the gluon coupling is found experimentally to agree with the SM prediction at a level of 1\%. Interpreting this bound within a one-partner model places a lower bound on fine tuning of about $1/35$ assuming $\Lambda=20$ TeV. The two-partner model {\it can} produce the gluon coupling within 1\% of the SM value with {\it smaller} fine-tuning; however, the regions of parameter space where this occurs (shaded in green in Fig.~\ref{fig:FTRg2}) are rather small, so an accidental cancellation is clearly involved. Note also that such accidental reduction in fine-tuning can only occur when $c_1$ and $c_2$ have opposite signs ($\theta<0$). The accidental nature of the fine-tuning reduction is further illustrated in Fig.~\ref{fig:FTtheta}: once the masses are fixed, fine-tuning drops significantly below the value inferred from the one-partner model only for a narrow range of the couplings. 

In principle, cancellation of the top quadratic divergence may involve $>2$ new particles, although we are not aware of any explicit model in which this is the case. It seems reasonable to conjecture that if this were the case, the correlation of Higgs coupling deviation and fine-tunings would persist, modulo possible accidental cancellations. 

\section{Conclusions}
\label{sec:concl}

In this paper, we pointed out and quantified a correlation between the level of fine-tuning of electroweak symmetry breaking and the deviations of the Higgs couplings to photons and gluons from their SM values. The connection holds in a very large class of 
well-motivated models: the basic assumptions are that the physics at the weak/TeV scale is weakly coupled, and that the quadratic divergence in the Higgs mass from the SM top loop is canceled by loops of new particles, the top partners. The top partners' contributions to the Higgs mass parameter and to the Higgs couplings to photons and gluons are determined by the same objects, their Higgs-dependent masses, resulting in a simple relationship between them. Thus, measuring Higgs couplings precisely provides a robust, model-independent test of naturalness. We showed that a measurement of Higgs couplings to gluons and photons at a per-cent level will either result in a discovery of a deviation from the SM, or imply that electroweak symmetry breaking is significantly tuned. This test of naturalness should be within the power of the proposed next-generation electron-positron collider such as the ILC. 

A potential ``loophole" in our argument is that the top partner contributions to the $hgg$ and $h\gamma\gamma$ couplings may be canceled by other non-SM contributions to these vertices. In the case of multiple top partners, there is also the possibility of cancellations of the top partners' contributions to $hgg$ and $h\gamma\gamma$ couplings among themselves. Typically, such cancellations should be regarded as accidental, and therefore unlikely. This was illustrated with an example of a two top partner model in Section~\ref{sec:2partners}. The only example that we are aware of where the cancellation of the top partner contributions to $hgg$ and $h\gamma\gamma$ happens for a reason that seems inherent to the structure of the theory and not accidental is the model studied in Ref.~\cite{Gilad}. However, the composite nature of the Higgs in that model implies large tree-level deviations of the Higgs couplings to $W$ and $Z$ bosons, and therefore it will still not escape detection via measurements of Higgs couplings. 

So far, naturalness of the electroweak scale has been mainly probed through direct searches for the top partners, which will of course continue in the next decade. We emphasize the complementarity between this program and the test of naturalness proposed here. The Higgs couplings test does not suffer from the well-known loopholes which plague direct searches ({\it e.g.} special spectra or R-parity violation). At the same time, there seems to be no reason for models where the deviations of $hgg$ and $h\gamma\gamma$ are suppressed, for whatever reason, to pose unusual difficulties for direct searches. Taken together, the two programs will provide a powerful and robust test of naturalness. 

In summary, we believe that the test of naturalness proposed here provides a compelling motivation for the future program of precision Higgs coupling measurements. We hope that this program will be realized in the coming years.

\vskip0.8cm
\noindent{\large \bf Acknowledgments} 
\vskip0.3cm

We would like to thank Christophe Grojean and Alfredo Urbano for useful discussions.
This research is supported by the U.S. National Science Foundation through grant PHY-0757868 and CAREER grant PHY-0844667.
NRL is supported by NSERC of Canada. MF would like to thank GGI for their hospitality during the completion of this work.

\vskip0.8cm
\noindent{\large \bf Note Added:} 
\vskip0.3cm

As we were completing this manuscript, Ref.~\cite{CEM} appeared on the {\tt arXiv.org} in which similar ideas were explored. 

%\section{Applicability to a Realistic Model: the MSSM Example}
%\draftnote{Put this one in an appendix?}


\begin{thebibliography}{99}

\bibitem{Peskin} 
  M.~E.~Peskin,
 {\it ``Comparison of LHC and ILC Capabilities for Higgs Boson Coupling Measurements,''}
  arXiv:1207.2516 [hep-ph].
  %%CITATION = ARXIV:1207.2516;%%

\bibitem{LRV} 
  I.~Low, R.~Rattazzi and A.~Vichi,
{\it ``Theoretical Constraints on the Higgs Effective Couplings,''}
  JHEP {\bf 1004}, 126 (2010)
  [arXiv:0907.5413 [hep-ph]].
  %%CITATION = ARXIV:0907.5413;%%

\bibitem{CFKV}
  D.~Carmi, A.~Falkowski, E.~Kuflik and T.~Volansky,
 {\it ``Interpreting LHC Higgs Results from Natural New Physics Perspective,''}
  JHEP {\bf 1207}, 136 (2012)
  [arXiv:1202.3144 [hep-ph]];\\
  %%CITATION = ARXIV:1202.3144;%%
  D.~Carmi, A.~Falkowski, E.~Kuflik, T.~Volansky and J.~Zupan,
 {\it ``Higgs After the Discovery: A Status Report,''}
  JHEP {\bf 1210}, 196 (2012)
  [arXiv:1207.1718 [hep-ph]].
  %%CITATION = ARXIV:1207.1718;%%
  
\bibitem{FGKM}
  S.~Fajfer, A.~Greljo, J.~F.~Kamenik and I.~Mustac,
 {\it ``Light Higgs and Vector-like Quarks without Prejudice,''}
  JHEP {\bf 1307}, 155 (2013)
  [arXiv:1304.4219 [hep-ph]].
  %%CITATION = ARXIV:1304.4219;%%  
  
\bibitem{LET} 
  J.~R.~Ellis, M.~K.~Gaillard and D.~V.~Nanopoulos,
 {\it ``A Phenomenological Profile of the Higgs Boson,''}
  Nucl.\ Phys.\ B {\bf 106}, 292 (1976);\\
  %%CITATION = NUPHA,B106,292;%%
  M.~A.~Shifman, A.~I.~Vainshtein, M.~B.~Voloshin and V.~I.~Zakharov,
 {\it ``Low-Energy Theorems for Higgs Boson Couplings to Photons,''}
  Sov.\ J.\ Nucl.\ Phys.\  {\bf 30}, 711 (1979)
  [Yad.\ Fiz.\  {\bf 30}, 1368 (1979)].
  %%CITATION = SJNCA,30,711;%%

\bibitem{NSUSYold}
  S.~Dimopoulos and G.~F.~Giudice,
 {\it ``Naturalness constraints in supersymmetric theories with nonuniversal soft terms,''}
  Phys.\ Lett.\ B {\bf 357}, 573 (1995)
  [hep-ph/9507282];\\
  %%CITATION = HEP-PH/9507282;%%
  A.~G.~Cohen, D.~B.~Kaplan and A.~E.~Nelson,
 {\it ``The More minimal supersymmetric standard model,''}
  Phys.\ Lett.\ B {\bf 388}, 588 (1996)
  [hep-ph/9607394].
  %%CITATION = HEP-PH/9607394;%%

\bibitem{GoldenSUSY}
 M.~Perelstein and C.~Spethmann,
 {\it ``A Collider signature of the supersymmetric golden region,''}
  JHEP {\bf 0704}, 070 (2007)
  [hep-ph/0702038].
  %%CITATION = HEP-PH/0702038;%%  
  
\bibitem{NSUSY}
  C.~Brust, A.~Katz, S.~Lawrence and R.~Sundrum,
 {\it ``SUSY, the Third Generation and the LHC,''}
  JHEP {\bf 1203}, 103 (2012)
  [arXiv:1110.6670 [hep-ph]];\\
  %%CITATION = ARXIV:1110.6670;%%  
  M.~Papucci, J.~T.~Ruderman and A.~Weiler,
 {\it ``Natural SUSY Endures,''}
  arXiv:1110.6926 [hep-ph];\\
  %%CITATION = ARXIV:1110.6926;%%
  Y.~Kats, P.~Meade, M.~Reece and D.~Shih,
{\it ``The Status of GMSB After 1/fb at the LHC,''}
  JHEP {\bf 1202}, 115 (2012)
  [arXiv:1110.6444 [hep-ph]];\\
  %%CITATION = ARXIV:1110.6444;%%
 N.~Desai and B.~Mukhopadhyaya,
 {\it ``Constraints on supersymmetry with light third family from LHC data,''}
  arXiv:1111.2830 [hep-ph].
  %%CITATION = ARXIV:1111.2830;%%

\bibitem{LH}
  N.~Arkani-Hamed, A.~G.~Cohen, E.~Katz and A.~E.~Nelson,
 {\it ``The Littlest Higgs,''}
  JHEP {\bf 0207}, 034 (2002)
  [hep-ph/0206021].
  %%CITATION = HEP-PH/0206021;%%

\bibitem{LHreviews}
For reviews and further references, see
  M.~Schmaltz and D.~Tucker-Smith,
 {\it ``Little Higgs review,''}
  Ann.\ Rev.\ Nucl.\ Part.\ Sci.\  {\bf 55}, 229 (2005)
  [hep-ph/0502182]; \\
  %%CITATION = HEP-PH/0502182;%%
  M.~Perelstein,
 {\it ``Little Higgs models and their phenomenology,''}
  Prog.\ Part.\ Nucl.\ Phys.\  {\bf 58}, 247 (2007)
  [hep-ph/0512128].
  %%CITATION = HEP-PH/0512128;%%        

\bibitem{5DHiggs}
  R.~Contino, Y.~Nomura and A.~Pomarol,
 {\it ``Higgs as a holographic pseudoGoldstone boson,''}
  Nucl.\ Phys.\ B {\bf 671}, 148 (2003)
  [hep-ph/0306259];\\
  %%CITATION = HEP-PH/0306259;%%
  K.~Agashe, R.~Contino and A.~Pomarol,
 {\it ``The Minimal composite Higgs model,''}
  Nucl.\ Phys.\ B {\bf 719}, 165 (2005)
  [hep-ph/0412089];\\
  %%CITATION = HEP-PH/0412089;%%
  R.~Contino, L.~Da Rold and A.~Pomarol,
 {\it ``Light custodians in natural composite Higgs models,''}
  Phys.\ Rev.\ D {\bf 75}, 055014 (2007)
  [hep-ph/0612048].
  %%CITATION = HEP-PH/0612048;%%  

\bibitem{spin1} 
  H.~Cai, H.~-C.~Cheng and J.~Terning,
 {\it ``A Spin-1 Top Quark Superpartner,''}
  Phys.\ Rev.\ Lett.\  {\bf 101}, 171805 (2008)
  [arXiv:0806.0386 [hep-ph]].
  %%CITATION = ARXIV:0806.0386;%%

\bibitem{TopShift} 
  T.~Han, H.~E.~Logan, B.~McElrath and L.~-T.~Wang,
 {\it ``Phenomenology of the little Higgs model,''}
  Phys.\ Rev.\ D {\bf 67}, 095004 (2003)
  [hep-ph/0301040];\\
  %%CITATION = HEP-PH/0301040;%%
    M.~Perelstein, M.~E.~Peskin and A.~Pierce,
 {\it ``Top quarks and electroweak symmetry breaking in little Higgs models,''}
  Phys.\ Rev.\ D {\bf 69}, 075002 (2004)
  [hep-ph/0310039];\\
  %%CITATION = HEP-PH/0310039;%%
  J.~Berger, J.~Hubisz and M.~Perelstein,
{\it ``A Fermionic Top Partner: Naturalness and the LHC,''}
  JHEP {\bf 1207}, 016 (2012)
  [arXiv:1205.0013 [hep-ph]].
  %%CITATION = ARXIV:1205.0013;%%

\bibitem{Gilad} 
  A.~Falkowski,
 {\it ``Pseudo-goldstone Higgs production via gluon fusion,''}
  Phys.\ Rev.\ D {\bf 77}, 055018 (2008)
  [arXiv:0711.0828 [hep-ph]];\\
  %%CITATION = ARXIV:0711.0828;%%
  A.~Azatov and J.~Galloway,
{\it ``Light Custodians and Higgs Physics in Composite Models,''}
  Phys.\ Rev.\ D {\bf 85}, 055013 (2012)
  [arXiv:1110.5646 [hep-ph]];\\
  %%CITATION = ARXIV:1110.5646;%%  
    CŽd.~Delaunay, C.~Grojean and G.~Perez,
{\it ``Modified Higgs Physics from Composite Light Flavors,''}
  arXiv:1303.5701 [hep-ph].
  %%CITATION = ARXIV:1303.5701;%%
  
\bibitem{Higgs_atlas}
  G.~Aad {\it et al.}  [ATLAS Collaboration],
{\it ``Search for the Standard Model Higgs boson in the diphoton decay channel with 4.9 fb$^{-1}$ of $pp$ collisions at $\sqrt{s}=7$ TeV with ATLAS,''}
  Phys.\ Rev.\ Lett.\  {\bf 108}, 111803 (2012)
  [arXiv:1202.1414 [hep-ex]]; see also ATLAS-CONF-2013-012;\\
  G.~Aad {\it et al.}  [ATLAS Collaboration],
 {\it ``Search for the Standard Model Higgs boson in the decay channel $H \to$ ZZ(*) $\to 4 \ell$ with 4.8 fb-1 of $pp$ collision data at $\sqrt{s}=7$ TeV with ATLAS,''}
  Phys.\ Lett.\ B {\bf 710}, 383 (2012)
  [arXiv:1202.1415 [hep-ex]]; see also ATLAS-CONF-2013-013;\\
  G.~Aad {\it et al.}  [ATLAS Collaboration],
 {\it ``Search for the Standard Model Higgs boson in the $H \to$ WW(*) $\to \ell \nu \ell \nu$ decay mode with 4.7 /fb of ATLAS data at $\sqrt{s}=7$ TeV,''}
  Phys.\ Lett.\ B {\bf 716}, 62 (2012)
  [arXiv:1206.0756 [hep-ex]]; see also ATLAS-CONF-2013-030;\\
  G.~Aad {\it et al.}  [ATLAS Collaboration],
{\it ``Search for the Standard Model Higgs boson in the $H$ to $\tau^{+} \tau^{-}$ decay mode in $\sqrt{s}=7$ TeV $pp$ collisions with ATLAS,''}
  JHEP {\bf 1209}, 070 (2012)
  [arXiv:1206.5971 [hep-ex]]; see also ATLAS-CONF-2012-160, ATLAS-CONF-2012-161.   
  
\bibitem{Higgs_cms}
S.~Chatrchyan {\it et al.} [CMS Collaboration], CMS-PAS-HIG-12-015; CMS-HIG-12-042; CMS-PAS-HIG-13-001; CMS-HIG-13-003; CMS-PAS-HIG-13-004; 
CMS-HIG-13-009; \\ 
S.~Chatrchyan {\it et al.}  [CMS Collaboration],
 {\it ``Search for the standard model Higgs boson decaying to bottom quarks in pp collisions at sqrt(s)=7 TeV,''}
  Phys.\ Lett.\ B {\bf 710}, 284 (2012)
  [arXiv:1202.4195 [hep-ex]].


\bibitem{Higgs_tev}
  T.~Aaltonen {\it et al.}  [CDF and D0 Collaborations],
 {\it ``Higgs Boson Studies at the Tevatron,''}
  arXiv:1303.6346 [hep-ex].
  %%CITATION = ARXIV:1303.6346;%%

\bibitem{Dittmaier:2011ti}
 S.~Dittmaier {\it et al.}  [LHC Higgs Cross Section Working Group Collaboration],
 {\it ``Handbook of LHC Higgs Cross Sections: 1. Inclusive Observables,''}
 arXiv:1101.0593 [hep-ph];\\
 %%CITATION = ARXIV:1101.0593;%%
 S.~Heinemeyer {\it et al.}  [ The LHC Higgs Cross Section Working Group Collaboration],
 {\it ``Handbook of LHC Higgs Cross Sections: 3. Higgs Properties,''}
 arXiv:1307.1347 [hep-ph].
 %%CITATION = ARXIV:1307.1347;%%
 
\bibitem{ATLAS:2013sla}
 [ATLAS Collaboration],
 {\it ``Combined coupling measurements of the Higgs-like boson with the ATLAS detector using up to 25 fb$^{-1}$ of proton-proton collision data,''}
 ATLAS-CONF-2013-034.
 %%CITATION = ATLAS-CONF-2013-034;%%
 
 \bibitem{Chatrchyan:2013lba}
 S.~Chatrchyan {\it et al.}  [CMS Collaboration],
 {\it ``Observation of a new boson with mass near 125 GeV in pp collisions at sqrt(s) = 7 and 8 TeV,''}
 JHEP {\bf 06}, 081 (2013)
 [arXiv:1303.4571 [hep-ex]].
 %%CITATION = ARXIV:1303.4571;%%

\bibitem{stopbounds}
S.~Chatrchyan {\it et al.}  [CMS Collaboration], {\it Search for top-squark pair production in the single lepton final state in pp collisions at 8 TeV}, CMS-PAS-SUS-13-011;\\
G.~Aad {\it et al.}  [ ATLAS Collaboration], {\it Search for direct production of the top squark in the all-hadronic ttbar + etmiss final state in 21 fb-1 of p-pcollisions at sqrt(s)=8 TeV with the ATLAS detector}, ATLAS-CONF-2013-024;\\
G.~Aad {\it et al.}  [ ATLAS Collaboration], {\it Search for direct top squark pair production in final states with one isolated lepton, jets, and missing transverse momentum in sqrts=8,TeV pp collisions using 21 fb?1 of ATLAS data}, ATLAS-CONF-2013-037.

\bibitem{stealthy}
  J.~Fan, M.~Reece and J.~T.~Ruderman,
{\it ``Stealth Supersymmetry,''}
  JHEP {\bf 1111}, 012 (2011)
  [arXiv:1105.5135 [hep-ph]].
  %%CITATION = ARXIV:1105.5135;%%
  
\bibitem{SUSY}
  R.~Dermisek and I.~Low,
{\it ``Probing the Stop Sector of the MSSM with the Higgs Boson at the LHC,''}
  Phys.\ Rev.\ D {\bf 77}, 035012 (2008)
  [hep-ph/0701235 [HEP-PH]];\\
  %%CITATION = HEP-PH/0701235;%%
  K.~Blum, R.~T.~D'Agnolo and J.~Fan,
 {\it ``Natural SUSY Predicts: Higgs Couplings,''}
  JHEP {\bf 1301}, 057 (2013)
  [arXiv:1206.5303 [hep-ph]];\\
  %%CITATION = ARXIV:1206.5303;%%
  R.~T.~D'Agnolo, E.~Kuflik and M.~Zanetti,
 {\it ``Fitting the Higgs to Natural SUSY,''}
  JHEP {\bf 1303}, 043 (2013)
  [arXiv:1212.1165 [hep-ph]];\\
  %%CITATION = ARXIV:1212.1165;%%
  G.~D.~Kribs, A.~Martin and A.~Menon,
  {\it ``Natural Supersymmetry and Implications for Higgs physics,''}
  arXiv:1305.1313 [hep-ph].
  %%CITATION = ARXIV:1305.1313;%%
      
\bibitem{CEM}
N.~Craig, C.~Englert and M.~McCullough,
{\it ``A New Probe of Naturalness,"}
arXiv:1305.5251 [hep-ph].

\end{thebibliography}
\end{document}